\documentclass[preprint]{aastex6}

\begin{document}


\title{Spatially-resolved spectroscopy of narrow-line Seyfert 1 host galaxies}


\author{J. Scharw\"achter\altaffilmark{1,2}, B. Husemann\altaffilmark{3,4}, G. Busch\altaffilmark{5}, S. Komossa\altaffilmark{6}, M. A. Dopita\altaffilmark{7}}
\altaffiltext{1}{Gemini Observatory, Northern Operations Center, 670 N. A'ohoku Place, Hilo, HI, 96720, USA}
\altaffiltext{2}{LERMA, Observatoire de Paris, PSL, CNRS, Sorbonne Universit\'es, UPMC, F-75014, Paris, France}
\email{jscharwaechter@gemini.edu}

\altaffiltext{3}{Max-Planck-Institut f\"ur Astronomie, K\"onigstuhl 17, 69117 Heidelberg, Germany}
\altaffiltext{4}{European Southern Observatory, Karl-Schwarzschild-Str. 2, 85748 Garching b. M\"unchen, Germany}

\altaffiltext{5}{I. Physikalisches Institut, Universit\"at zu K\"oln, Z\"ulpicher Str. 77, 50937 K\"oln, Germany}

\altaffiltext{6}{National Astronomical Observatories, Chinese Academy of Sciences, Beijing 100012, China}

\altaffiltext{7}{RSAA, The Australian National University, Cotter Road, Weston Creek, ACT 2611 Canberra, Australia}


\begin{abstract}
We present optical integral field spectroscopy for five $z<0.062$ narrow-line Seyfert 1 galaxies (NLS1s) host galaxies, probing their host galaxies at $\gtrsim 2-3$~kpc scales. Emission lines in 
the nuclear AGN spectra and the large-scale host galaxy are analyzed separately, based on an AGN-host decomposition technique. The host galaxy gas kinematics indicates large-scale gas rotation in all five sources.
At the probed scales of $\gtrsim 2-3$~kpc, the host galaxy gas is found to be predominantly ionized by star formation without any evidence of a strong AGN contribution. None of the five objects shows specific star formation rates exceeding the main sequence of low-redshift star forming galaxies. The specific star formation rates for \object{MCG-05-01-013} and \object{WPVS 007} are roughly consistent with the main sequence, while \object{ESO~399-IG20}, \object{MS 22549-3712}, and \object{TON S180} show lower specific star formation rates, intermediate to the main sequence and red quiescent galaxies. The host galaxy metallicities, derived for the two sources with sufficient data quality (\object{ESO~399-IG20} and \object{MCG-05-01-013}), indicate central oxygen abundances just below the low-redshift mass-metallicity relation. Based on this initial case study, we outline a comparison of AGN and host galaxy parameters as a starting point for future extended NLS1 studies with similar methods.
\end{abstract}

\keywords{galaxies: active --- galaxies: individual (ESO~399-IG20, MCG-05-01-013, MS 22549-3712, WPVS 007, TON S180)}



\section{Introduction} \label{sec:intro}

The continuum emission associated with accreting black holes (BHs) is a major source of ionizing photons, causing a rich emission-line 
 spectrum in the surrounding gas. The prominent emission-line spectra associated with AGN have been used as an
 observational tool to study the properties of gas in the vicinity of an AGN.
 Principal component analysis has revealed that the optical emission-line properties of AGN are dominated by 
 two correlations \citep[e.g.][]{Boroson:1992aa,Boroson:2002aa}:
The first principal component (Eigenvector 1) mainly projects onto the strengths of \ion{Fe}{2} and [\ion{O}{3}] (which are anticorrelated) and to a lesser extent onto the
asymmetry and line width of H$\beta$; the second principal component (Eigenvector 2) mainly projects onto AGN
luminosity, which is anticorrelated with the
strength of \ion{He}{2}$\lambda4686$. 
The correlations are likely to hold important clues on the main parameters that drive nuclear activity.
Eddington ratio and accretion rate have been suggested as the primary drivers of the first and second principal component, respectively, 
\citep{Boroson:2002aa, Shen:2014aa}, while the narrow-line region (NLR) density may also be relevant for the first principal component \citep{Xu:2012aa}. 

NLS1s are likely to provide further insights into these correlations, since they occupy one extreme end
in the parameter space \citep[e.g. review by][]{Komossa:2008aa}. 
NLS1s are characterized by a narrow width of the broad permitted lines ($\mathrm{FWHM(H\beta)} < 2000\ \mathrm{km\ s^{-1}}$), strong \ion{Fe}{2} emission, and flux ratios of [\ion{O}{3}]$\lambda 5007$/$\mathrm{H\beta} < 3$ typical of Seyfert~2 galaxies \citep{Shuder:1981aa, Osterbrock:1985aa, Goodrich:1989aa}. The classical H$\beta$ width criterion $\mathrm{FWHM(H\beta)} < 2000\ \mathrm{km\ s^{-1}}$ has been 
controversial, since NLS1-like properties are also found in objects showing wider line widths \citep[e.g.][]{Sulentic:2000aa,Veron-Cetty:2001aa}. Furthermore, it is not uncommon to find NLS1s with much broader underlying Gaussian components, similar to the broad-line components of broad-line Seyfert 1s \citep[e.g.][]{Mullaney:2008aa}. This has led to the view that NLS1s are not a distinct ``class'' of AGN, but rather represent a smooth transition toward the more extreme end of the
AGN parameter space.

The 
[\ion{O}{3}] lines of NLS1s often show complex line profiles and blueshifts, indicating radiation-pressure-driven winds and outflows in the nuclear region \citep[e.g.][]{Marziani:2003aa,Bian:2005aa,Komossa:2008ab}. In the X-rays, NLS1s show, on average, steeper X-ray slopes and higher variability than their broad-line counterparts \citep{Boller:1996aa}. There are indications for lower electron densities in the NLRs of NLS1s compared to broad-line Seyfert 1 galaxies (BLS1s), \citep{Xu:2007aa, Rakshit:2017aa}, though this trend is less clear than other NLS1 properties.

NLS1s are considered as local candidates for rapid BH growth and strong radiation-pressure-driven feedback.
The extreme properties of NLS1s suggest high accretion rates onto black holes of relatively low mass \citep[e.g.][]{Boroson:2002aa,Grupe:2004aa},
which could be consistent with young AGN experiencing a phase of rapid BH growth.
The location of NLS1s in the BH scaling relations has been a matter of controversy. Some studies place NLS1s onto the $M_\mathrm{BH}$-$\sigma_\star$ relation, 
while others suggest that they lie below the relation and may be in the process of growing their BHs towards the relation \citep[e.g.][]{Komossa:2007aa, Mathur:2012aa, Woo:2015aa, Rakshit:2017aa, Zhou:2006aa}.
The host galaxies of 
NLS1s seem to be dominated by secular evolution and show indications of pseudo-bulges  
\citep{Orban-de-Xivry:2011aa,Mathur:2012aa}. They may thus be similar to inactive galaxies with pseudo-bulges, which have been suggested 
to be inconsistent with the BH scaling relations \citep{Kormendy:2011aa}. 

The number of NLS1s in the nearby Universe is limited. Only 35 out of the 2011 NLS1s found in the Sloan Digital Sky Survey (SDSS)
Data Release 3 by \citet{Zhou:2006aa} are located at $z<0.06$. Data Release 12 has recently led to a factor of five larger NLS1 catalog \citep{Rakshit:2017aa}, of which about
100 objects are found at $z < 0.06$. 
Since the majority of NLS1s are found at higher redshift, much of the current knowledge on NLS1s has been derived from X-ray observations or single spectra. 
Among the nearby NLS1s at $z\lesssim 0.06$ (see samples by \citet{Veron-Cetty:2001aa} or \citet{Sani:2010aa}), 
only very few objects have literature data based on optical or near-infrared integral field spectroscopy. These few exceptions include 
\object{NGC~4051}, \object{Mrk~766}, \object{Mrk 493}, and \object{Mrk 1044}, \citep{Sosa-Brito:2001aa,Barbosa:2006aa,Riffel:2008aa,Popovic:2009aa,Schonell:2014aa,Dopita:2015aa}.
The more detailed case studies among these have focussed on near-infrared emission lines in \object{NGC~4051} \citep{Riffel:2008aa}, revealing a nuclear H$_2$ molecular gas inflow, and 
 \object{Mrk~766} \citep{Schonell:2014aa}, where the H$_2$ emission is found to be dominated by rotation, while the [\ion{Fe}{2}] emission shows signs of an outflowing component.
At slightly higher redshift ($\sim 0.03 <z<0.3$), a few sources with NLS1 properties are covered by the optical IFU studies by \citet{Husemann:2008aa} and \citet{Husemann:2014aa}. However, the spatial resolution is already more limited in this redshift range.

In this paper, we focus on the largely unexplored domain of spatially-resolved optical spectroscopy for nearby NLS1s and present an initial case study of five $z < 0.062$ AGN with NLS1 characteristics, including \object{ESO~399-IG20}, \object{MCG-05-01-013}, \object{MS 22549-3712}, \object{TON S180}, and \object{WPVS 007} (see Table~\ref{table:galprop}). Based on available literature data, this small inhomogeneous sample covers 
a variety of AGN parameters from more prototypical NLS1s to  NLS1/BLS1 borderline
objects. All five sources were classified as ``S1n'' (i.e. broad Balmer line components 
narrower than $2000\ \mathrm{km\ s^{-1}}$) in the catalog by \citet{Veron-Cetty:2010aa}. \object{MS 22549-3712}, \object{WPVS 007}, and \object{TON S180} are rather well-known NLS1s
and have previously been included in NLS1-focussed samples
\citep[e.g.][]{Veron-Cetty:2001aa, Leighly:2009aa, Mathur:2012aa}.
\object{ESO~399-IG20} and \object{MCG-05-01-013} are less well-studied.
\citet{Dietrich:2005aa} classified \object{ESO~399-IG20} as a NLS1, though it had previously been associated with BLS1s
(see \citet{Panessa:2011aa} and references therein). \object{MCG-05-01-013} was included in the sample of 
soft X-ray selected AGN by \citet{Grupe:2004ab}. However, with $\mathrm{FWHM(H\beta)=2400\ km\ s^{-1}}$, the width of the broad 
H$\beta$ component was found to be just above the
classical NLS1 criterion. 

Based on new optical IFU data and an AGN-host deblending technique, 
we review the nuclear parameters of these sources and discuss excitation mechanisms, gas kinematics, star formation rates, 
and gas metallicity in the host galaxies. The data reduction and analysis techniques, including the AGN-host deblending, 
stellar continuum subtraction, and emission-line fitting, 
are explained in Section~\ref{sec:obs}. The resulting AGN parameters and host galaxy properties are presented in Sections~\ref{sec:obs:nuclfit} and \ref{sec:host}, 
respectively. 
In Section~\ref{sec:conclusion}, we summarize and discuss the results in the context of the wider literature on NLS1s.
Throughout the paper we assume $H_0 = 70\ \mathrm{km\ s^{-1}\ Mpc^{-1}}$, $\Omega_M=0.3$, and $\Omega_\Lambda = 0.7$.

\floattable
\begin{deluxetable}{lccc}
\tablecaption{Basic properties of the five NLS1s \label{table:galprop}}
\tablecolumns{5}
\tablewidth{0pt}
\tablehead{
\colhead{Object} &
\colhead{z} & \colhead{$\mathrm{D_L}$} & \colhead{Scale} \\
\colhead{} &  \colhead{} &
\colhead{(Mpc)} & \colhead{(pc arcsec$^{-1}$)}
}
\decimalcolnumbers
\startdata
\object{ESO~399-IG20}  & 0.025\tablenotemark{a} & 106 & 491  \\
\object{MCG-05-01-013}  & 0.031\tablenotemark{b} & 130 & 595\\
\object{MS 22549-3712}  & 0.039\tablenotemark{c}  & 168 & 756\\
\object{TON S180}  & 0.062\tablenotemark{d}  & 273 & 1178\\ 
\object{WPVS 007}  & 0.029\tablenotemark{e} & 123 & 562 \\
\enddata
\tablecomments{Column 1: Object name; Column 2: Redshift z (taken from \tablenotemark{a}\citet{Dietrich:2005aa}, \tablenotemark{b}{\citet{Domingue:2000aa}}, \tablenotemark{c}{\citet{Grupe:1998aa}}, 
\tablenotemark{d}{\citet{Wisotzki:1995aa}}, \tablenotemark{e}{\citet{Jones:2010aa}} and rounded to the third decimal place); 
Columns 3 and 4: Luminosity distance $D_L$ and scale, based on \citet{Wright:2006aa} and $H_0 = 70\ \mathrm{km\ s^{-1}\ Mpc^{-1}}$, $\Omega_M=0.3$, and $\Omega_\Lambda = 0.7$ after correction to the reference frame of the cosmic microwave background \citep{Fixsen:1996aa}, (see NASA/IPAC Extragalactic Database (NED) at {\it https://ned.ipac.caltech.edu/}).}

\end{deluxetable}

\section{Observations and data reduction}\label{sec:obs}

The five sources were observed between 2011 September 30 and October 4
using the Wide Field Spectrograph WiFeS \citep{Dopita:2007aa, Dopita:2010aa}, (ANU 2.3~m telescope
at Siding Spring, Australia).
WiFeS provides optical integral field spectroscopy over a field-of-view of 25\arcsec\ $\times$ 38\arcsec, 
composed of 25 slitlets of 1\arcsec\ $\times$ 38\arcsec\ in image-slicer design. Simultaneous observations in a blue and red arm
result in a wide wavelength coverage. 
In the blue arm, all sources were observed using the $R_S = 3000$ grating ($B3000$). In the red arm, the 
$R_S = 7000$ grating ($R7000$) was used for ESO 399-IG20, MCG-05-01-013,
MS 22549-3712, and WPVS 007. For TON S180, the $R_S = 3000$ grating ($R3000$) was chosen in the red, because at $z=0.062$
the [\ion{S}{2}]$\lambda\lambda$6717,6731 lines are shifted outside the
wavelength range of the $R7000$ grating. 
The total integration time for each object was split into three or more individual exposures for improved
cosmic ray removal. The observing conditions were variable and largely non-photometric.
Table~\ref{table:obslog} shows a summary of the observational setups for the
five sources, together with the resulting image quality.

\floattable
\begin{deluxetable}{lccccc}
\tablecaption{Log of observations \label{table:obslog}}
\tablecolumns{7}
\tablewidth{0pt}
\tablehead{
\colhead{Object} & \colhead{Observing date} & \colhead{Integration time} & 
\colhead{Grating} & \colhead{FHWM near H$\beta$}  &  \colhead{FHWM near H$\alpha$}\\
\colhead{} & \colhead{} & \colhead{(s)} & 
\colhead{(blue/red)} & \colhead{(arcsec)} & \colhead{(arcsec)}
}
\decimalcolnumbers
\startdata
 \object{ESO~399-IG20} & 2011-09-30 & 1800  & B3000/R7000 & 4.2 & 3.4 \\      
   \object{MCG-05-01-013} & 2011-10-02 & 2400  & B3000/R7000  & 3.3 & 3.2 \\
  \object{MS 22549-3712} & 2011-10-03 & 5400   & B3000/R7000 & 2.8\tablenotemark{a} & 2.6\tablenotemark{a} \\
    \object{TON S180} & 2011-10-04 & 3000   & B3000/R3000 & 2.7\tablenotemark{a} & 2.7\tablenotemark{a} \\ 
    \object{WPVS 007} & 2011-09-30 & 3000 & B3000/R7000  & 3.3\tablenotemark{a} & 3.1\tablenotemark{a}\\
\enddata
\tablenotetext{a}{For the emission line analysis in this paper, the data for 
\object{MS 22549-3712}, \object{WPVS 007}, and \object{TON S180} were binned to a sampling of 3\arcsec\ per pixel. The FWHM was measured based on the original
pixel sampling of 1\arcsec\ per pixel.}
\tablecomments{Column 1: Object name; Column 2: Observing date; Column 3: Total on-source integration time; 
Column 4: Grating combination used for the WiFeS blue and red arms; Columns 5 and 6: FWHM of the PSFs corresponding to the wavelength regions 
around H$\beta$ and H$\alpha$, respectively. The PSFs were obtained as described in Sect.~\ref{sec:deblend}. We report the direct FWHMs measured via IRAF tasks \textsc{IMEXAMINE} or
\textsc{PSFMEASURE}, which are very similar to the FWHMs derived from a Moffat profile fit.}
\end{deluxetable}

The data were reduced using the reduction recipes implemented in the 
python WiFeS pipeline {\sc PyWiFeS} version 0.6.0 \citep{Childress:2013aa}, largely following standard procedures.
All data were observed in nod-and-shuffle mode with exposures alternating between the object and a nearby sky field used
for sky subtraction.
The basic calibrations are based on 
day-time dome flat fields, night-time Ne-Ar arc lamp exposures for wavelength calibration, and
day-time flat-lamp exposures of a wire coronograph for spatial rectification. Because of adverse
weather conditions, no twilight flats could be taken. Single night-time bias frames were
observed close in time to the science data in order to improve the subtraction of the curved and temporally variable
bias level. During the reduction, a model of the corresponding bias frame --
obtained by median-averaging each row in each of the four quadrants -- was subtracted from each science frame
after an initial overscan-subtraction.
The data were flux calibrated and corrected for telluric absorption features, using observations of 
spectrophotometric standard stars and early-type $B$V stars, respectively.  
All object exposures were reduced separately and median-combined for the final data cube which has a
native sampling of 1\arcsec\ per pixel.
Table~\ref{table:obslog} shows the spatial resolution achieved for these objects in the
native pixel sampling.

Due to the variable and largely non-photometric conditions, the accuracy of the absolute flux calibration based on the spectrophotometric standard stars is limited.
 By comparing published broad-band fluxes with the ones derived from the WiFeS data 
 and by comparing the sensitivity curves from the different observing nights, the uncertainty in the absolute flux calibration is conservatively estimated to be less than a factor of 2 (i.e. 0.75~mag). This corresponds to an uncertainty of $\lesssim 0.3$~dex for the logarithmic specific star formation rates reported below and to an uncertainty of a factor of $\lesssim \sqrt{2}$ for BH masses and Eddington ratios.  
 
\newpage

\subsection{AGN-host deblending}\label{sec:deblend}

In order to study the host galaxies underneath the bright point-like AGN emission, we decomposed the data cubes into AGN and host contributions using the
software package \textsc{QDeblend$^{\mathrm{3D}}$} \citep{Husemann:2013aa,Husemann:2014aa,Husemann:2016aa}. 
A detailed description of the deblending algorithm can be 
found in the above articles. The basic method relies on the fact that 
the spatial 
distribution of the broad-line flux in the data cube provides a measure of the point-spread function (PSF) 
at the central broad-line wavelength, given that
the broad 
Balmer lines of unobscured AGN originate from an unresolved region 
$<1\,\mathrm{pc}$. 

Assuming that the central-most 
spectrum in the data cube is dominated by the AGN, the AGN contribution across the IFU field-of-view is removed by 
normalizing the AGN spectrum in absolute flux according to the PSF at a given spaxel. Depending 
on the AGN-host galaxy contrast, even the central spectrum can have a significant host galaxy contribution.
Therefore, the host-galaxy contribution is iteratively removed from the AGN 
spectrum while preserving the host-galaxy surface-brightness distribution extrapolated 
from the circumnuclear spectra (after first iteration) to the centre for each wavelength slice. 

The good spectral resolution of 
WiFeS allows us to separate the broad lines from the narrow lines for a PSF measurement, even though the broad lines of NLS1s are, by 
definition, narrower than those of BLS1s. 
As the PSF changes with wavelength due to atmospheric dispersion, the cubes obtained from the WiFeS blue and red arms were treated separately during the deblending process, 
using the PSFs derived from the broad lines of
H$\beta$ and H$\alpha$, respectively.

\begin{figure*}
\figurenum{1}
\plotone{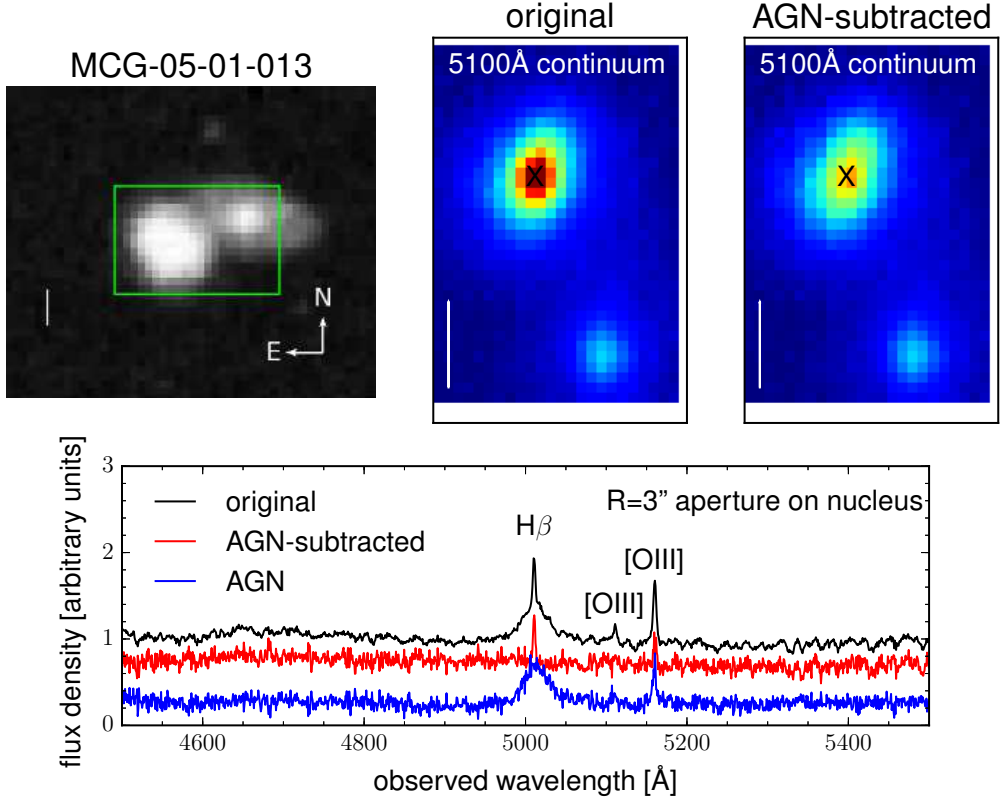}
\caption{Illustration of the AGN deblending for the example of \object{MCG-05-01-013}. Top left panel: DSS image with an overlay of the WiFeS field-of-view. 
  Top middle and right panels: Comparison of the rest-frame 5100~\AA\ continuum
  images of the original WiFeS data cube, containing AGN and host galaxy emission, and the cube after AGN subtraction. The continuum images were
  obtained by spectrally averaging over the observed wavelength range between 5240 and 5270~\AA. East is up, north is to the right. The scale bars indicate 5~kpc. 
  The black cross indicates the position of the AGN, as determined from an elliptical Gaussian fit to the host-cleaned AGN cube
    integrated over the [\ion{O}{3}]$\lambda 5007$ line. 
  Bottom panel: Comparison of the nuclear spectra from the original
  WiFeS data cube (black), the AGN-subtracted one (red), and the AGN cube (blue). 
  The spectra are integrated over a 3\arcsec-radius pseudo-circular aperture centered on the nucleus. 
  \label{fig:qdebl}}
\end{figure*}

The morphology and surface brightness profile for most of the targeted AGN host galaxies have not been studied in detail before. Therefore,
we obtained a model of the host-galaxy
surface-brightness distributions by fitting the  
synthetic $g$ and $r$ broad-band images extracted from the WiFeS data cubes. The fits were derived using 
\textsc{GALFIT} \citep{Peng:2002aa,Peng:2010aa} and assuming a 
S\'{e}rsic profile for the host galaxies (as well as for a companion galaxy, where applicable).
For the convolution kernel, we used the PSFs estimated via \textsc{QDeblend$^{\mathrm{3D}}$} from the Balmer line closest in wavelength. 
Bad spaxels and foreground stars were masked out during the fitting process. We explored two simple 
models with S\'{e}rsic indices of $n=1$ and $n=4$, respectively, and selected the model with the lower 
chi-square to derive the magnitudes of the AGN and the host galaxy as well as the effective 
radius.
 Fig.~\ref{fig:qdebl} shows an example of a spectrum in the 
IFU data cube before and after AGN decomposition, which demonstrates that the AGN contribution is efficiently removed by the deblending
procedure.

\newpage

\subsection{Continuum subtraction and emission-line fitting for the host galaxies}
The underlying host galaxy spectrum in the AGN-deblended data -- consisting of a stellar continuum and possibly emission lines -- was 
fitted using the spectral synthesis model code \textsc{PyParadise}, which is an extended Python version of \textsc{paradise} \citep{Walcher:2015aa}.
\textsc{PyParadise} takes a set of template continuum spectra to reconstruct the best-fitting continuum shape and a 
Gaussian profile for each emission line. The modeling process is briefly outlined in the following. Further details 
of the software can be found in \citet{Walcher:2015aa}. 

In a first step, all library spectra and all spectra in the data cube were normalized by a running mean filter, 
where regions of strong emission-line or sky-line residuals were replaced by a linear 
interpolation over the masked wavelength ranges. Then the template library was smoothed to the 
instrumental resolution of the target spectra. A single spectrum of the library was chosen in order to
estimate the systematic redshift and velocity dispersion by convolving the spectrum with a corresponding kernel as 
part of an MCMC algorithm. The entire library was then convolved with the maximum 
likelihood kinematics, and a non-linear negative least square minimization was used to find the best linear 
combination of the library spectra that matches the data. In three subsequent iterations, the best 
template combination was again used to refine the kinematics. After the best-fitting 
stellar continuum model was removed from the real data, the emission lines were fitted in the residual spectra. Here,
all lines were modeled with simple Gaussian profiles and coupled in redshift and intrinsic rest-frame velocity dispersion while 
taking into account variations in spectral resolution.

A Monte-Carlo bootstrap was used to estimate errors on the emission lines, including template mismatches of the stellar population spectra library. 
For each spectrum, the fitting process 
giving the best-fitting stellar kinematics was repeated 
200 times after modulating the spectra according to the flux noise per spectral pixel and 
considering just 80\% of the library spectra randomly chosen for each 
realization. Likewise, the emission-line fits were repeated in order to measure errors for the line redshift, velocity dispersion, and line flux 
based on the rms of the results. 
The entire process -- except for the initial smoothing of the template library -- was performed on a spaxel-by-spaxel basis, providing maps 
for the stellar kinematics, the line fluxes, as well as the gas kinematics 
and velocity dispersion across the field-of-view. In order to improve the signal-to-noise ratio,
the data cubes for \object{MS 22549-3712}, \object{WPVS 007}, and \object{TON S180} were binned to
3\arcsec\ per pixel.

\subsection{Emission-line fitting for the nuclear AGN spectra}\label{sec:AGNfit}
The AGN emission lines of the five sources were measured in the nuclear spectra
obtained by integrating the host-deblended AGN
data cubes over the full PSF. The emission lines were fitted 
simultaneously in two wavelength
regions around H$\beta$ and H$\alpha$ using a superposition of Gaussian profiles 
 together with \ion{Fe}{2} templates from \citet{Kovacevic:2010aa}. 
The underlying AGN continuum was sampled in the line-free regions of the spectra and subtracted using a first-order 
polynomial fit for each of the two wavelength regions.

\begin{figure*}
\figurenum{2}
\gridline{\fig{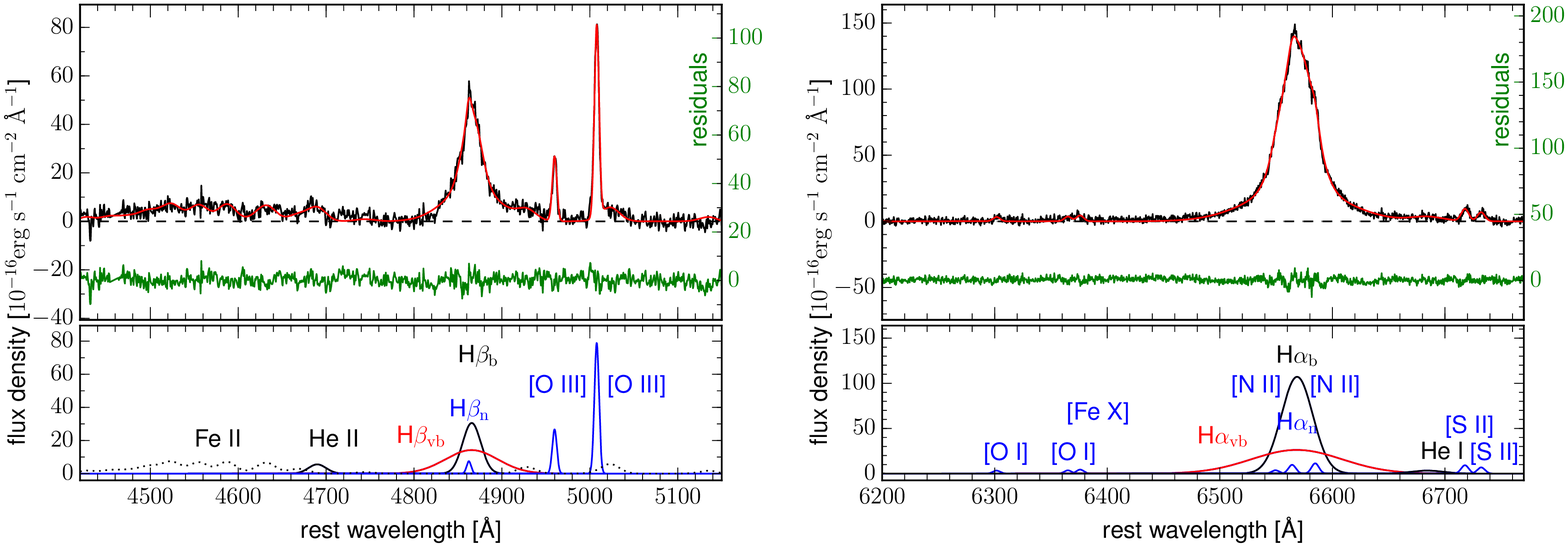}{\textwidth}{(a) ESO~399-IG20}
          }
\gridline{\fig{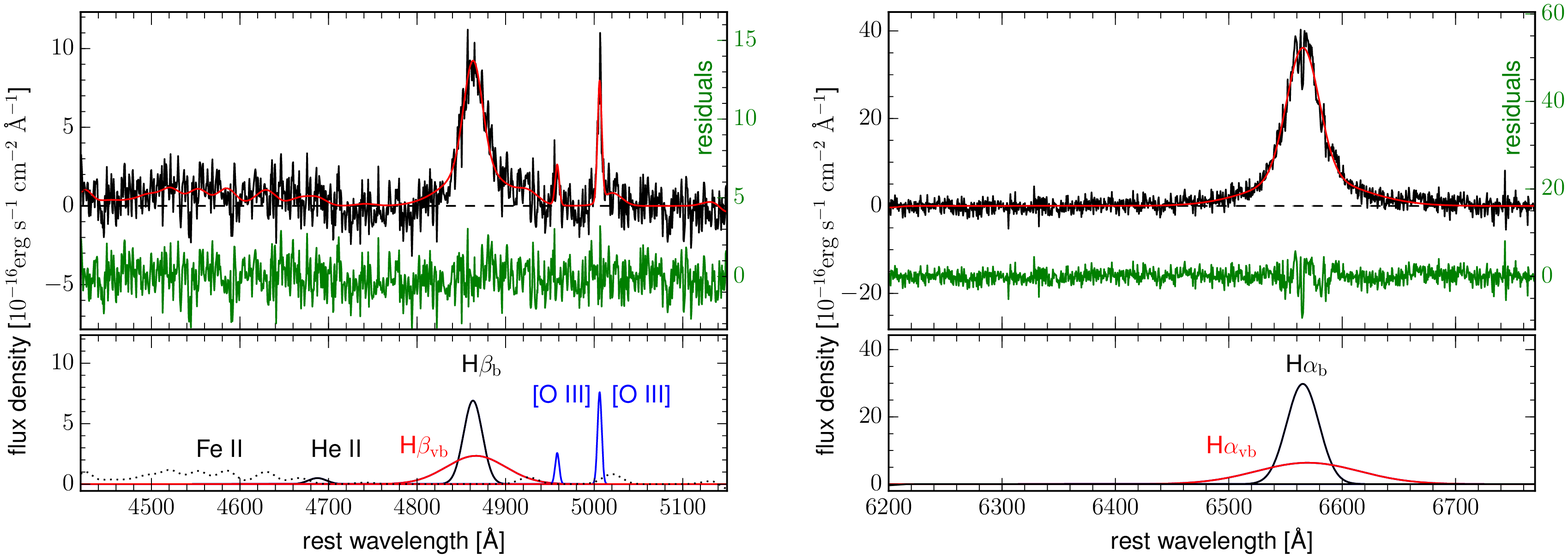}{\textwidth}{(b) MCG-05-01-013}
          }
\gridline{\fig{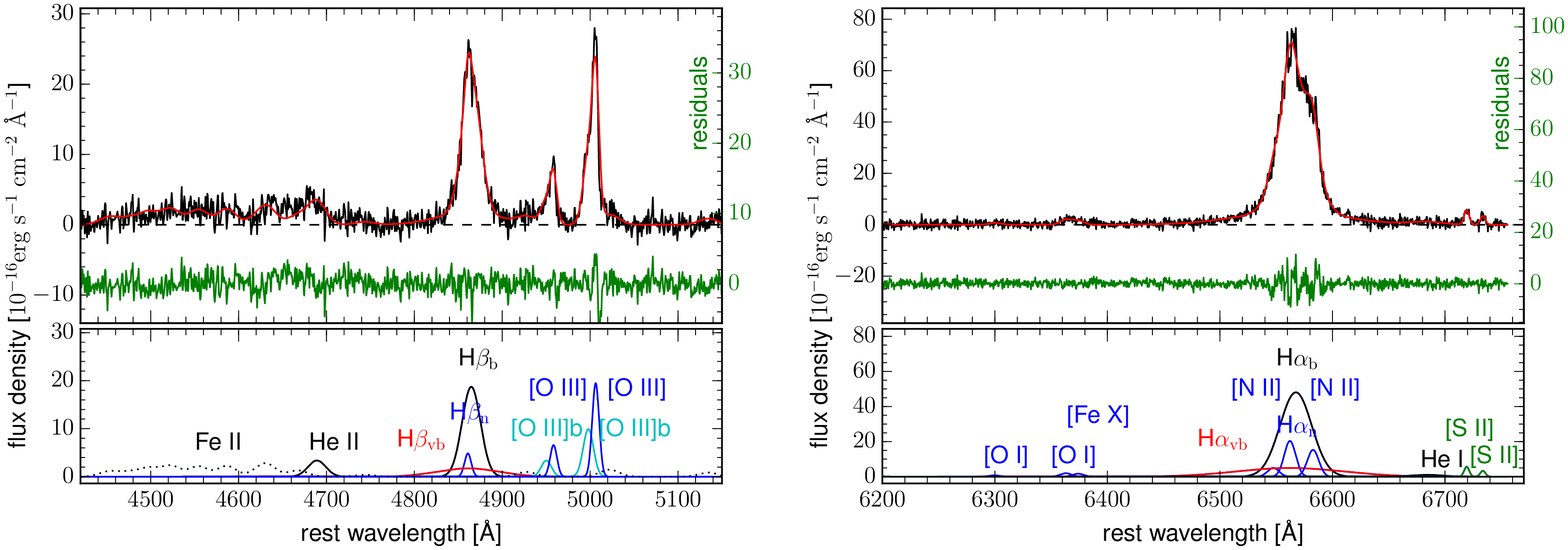}{\textwidth}{(c) MS 22549-3712}
          }
\end{figure*}
\begin{figure*}
\figurenum{2}
\gridline{\fig{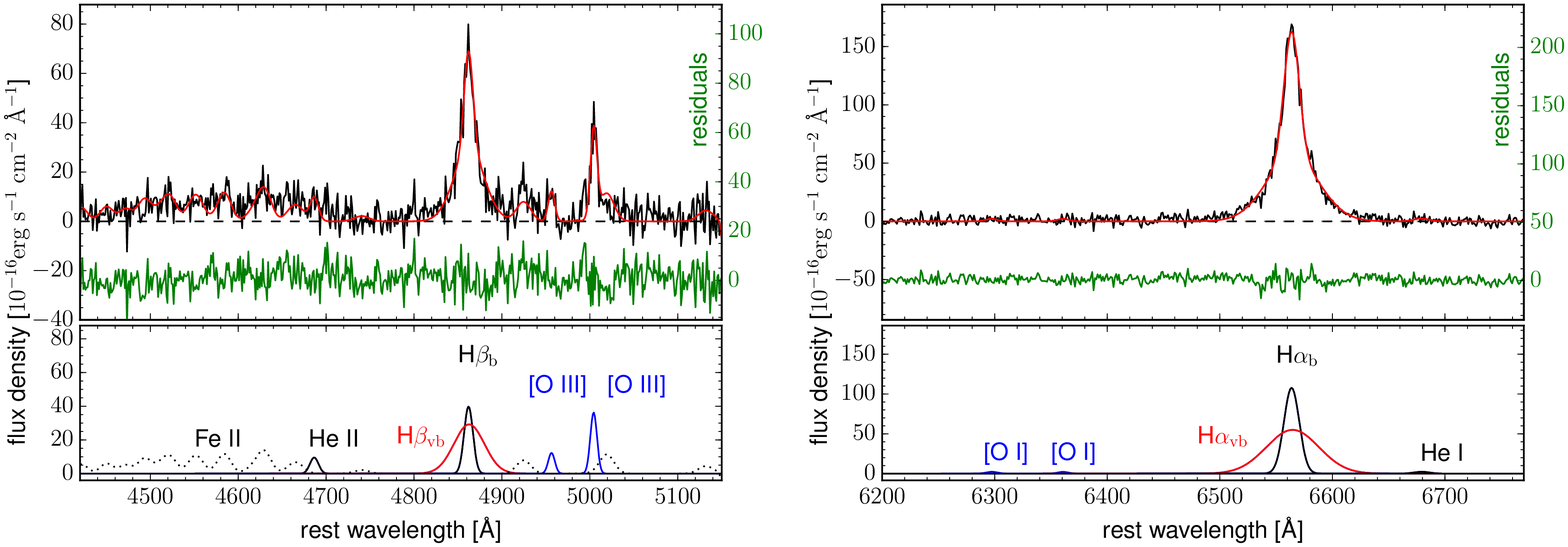}{\textwidth}{(d) TON S180}
          }
\gridline{\fig{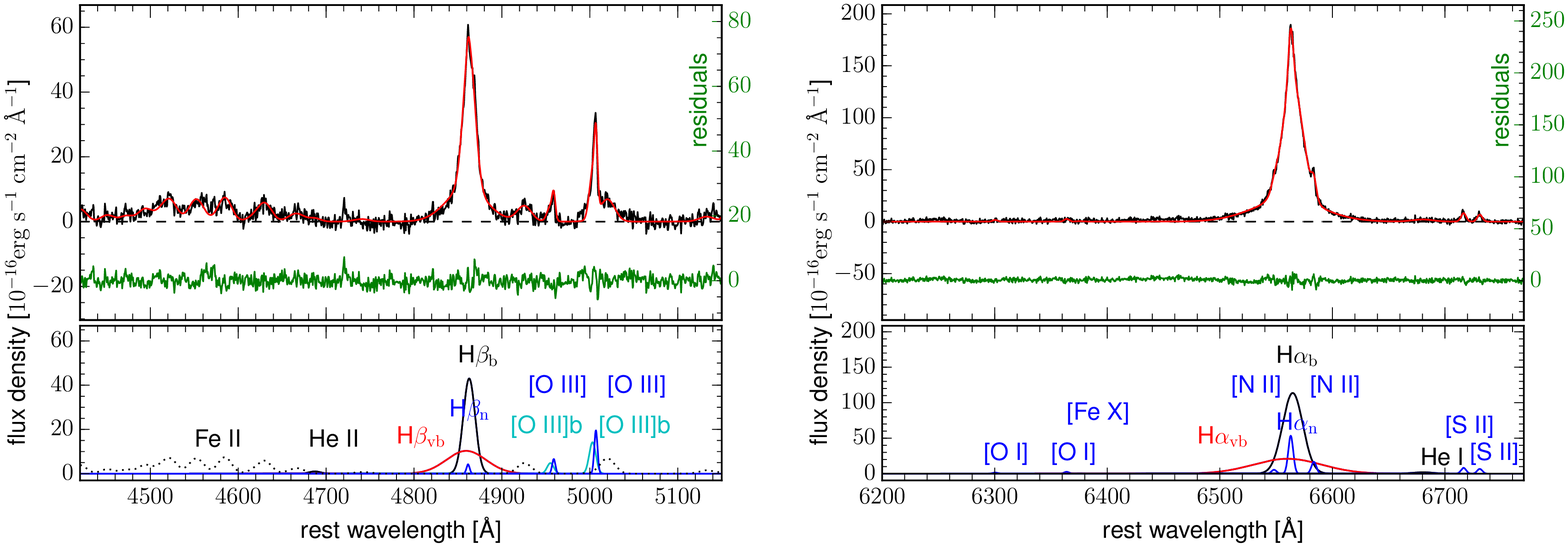}{\textwidth}{(e) WPVS 007}
          }
\caption{Fit to the nuclear spectra of the NLS1s after continuum subtraction. 
The spectra are fitted in the wavelength ranges around H$\beta$ (left panels) and
H$\alpha$ (right panels). Both wavelength ranges were fitted simultaneously, as explained in the text. 
The upper panels show the spectrum (black), the best-fitting model (red) and the residuals (green). 
The lower panels show the individual line components of the best-fitting model. Lines showing the same color were kinematically tied to each other during the fit.  
\ion{Fe}{2} is marked by a dotted line for clarity, but was kinematically tied to the other components shown in black.
\label{fig:nuclfit}}
\end{figure*}

The emission-line fits for the continuum-subtracted AGN spectra 
are shown in Fig.~\ref{fig:nuclfit}. 
The broad components of H$\beta$ and H$\alpha$ were modeled with a superposition of two Gaussian profiles (labeled ``broad'' (b) and 
``very broad'' (vb) component, respectively). 
For consistency, the very broad Gaussian component was included for all sources in order to improve           
the fit to the broad line wings, which were otherwise underrepresented.                                                           
Our modeling approach for the broad H$\beta$ and H$\alpha$ components is in line
with a larger number of previous studies which conclude that the broad H$\beta$ and H$\alpha$ components in many NLS1s
are best approximated via two Gaussian components or a Lorentzian profile
\citep[e.g.][]{Dietrich:2005aa, Xu:2007aa, Mullaney:2008aa, Schmidt:2016aa,Cracco:2016aa}. 

During the fit, the
corresponding components of H$\beta$ and H$\alpha$ were kinematically
tied to each other in terms of central velocity and line width. The kinematics of \ion{He}{2} and \ion{He}{1} was tied to the
``broad'' Gaussian component.
The line profiles of all 
other lines -- including the narrow components of H$\beta$ and H$\alpha$ -- are based on a single narrow Gaussian profile. 
In order to improve the convergence of
the fit and the decomposition of the complex H$\beta$ and H$\alpha$ profiles, 
all narrow components were forced to have the same relative velocity and
line width. 
An additional blue-shifted component of the [\ion{O}{3}]$\lambda\lambda 4959$,5007 lines was included in the models
when the [\ion{O}{3}] lines showed obvious asymmetries. 
As a further
constraint for all fits, we imposed constant flux ratios of 3.05 and 2.948 for the doublet lines [\ion{N}{2}]$\lambda\lambda 6581, 6548$
 and [\ion{O}{3}]$\lambda\lambda 5007, 4959$ (main as well as blueshifted components), respectively
 \citep[e.g.][]{Schirmer:2013aa}.

The kinematics of the \ion{Fe}{2} templates
was tied to the ``broad'' Gaussian component of the H$\beta$ and H$\alpha$ model. 
We used the narrowest \ion{Fe}{2} templates
provided by \citet{Kovacevic:2010aa} and convolved the templates to the corresponding Gaussian width using a Gaussian kernel. The
instrumental spectral resolution of the WiFeS observations was taken into account during the convolution. 
The intrinsic width of the available \ion{Fe}{2} templates of $\sim 500\ \mathrm{km\ s^{-1}}$ (Gaussian
sigma) imposed a lower limit on the possible \ion{Fe}{2} widths probed by the fitting procedure. 
For \object{WPVS 007} and \object{TON S180}, the width of the ``broad'' Gaussian reference component was below this limit ($\sim 412\ \mathrm{km\ s^{-1}}$ and $\sim 306\ \mathrm{km\ s^{-1}}$, respectively), so that the \ion{Fe}{2} width was effectively decoupled from the reference component.

\section{Results and discussion}\label{sec:results}

\subsection{AGN parameters}\label{sec:obs:nuclfit}
The line fluxes and kinematics derived from the fit to the host-deblended AGN spectra of the five sources are listed in 
Table~\ref{table:fluxes}.
The narrow lines in all five sources are characterized by velocity dispersions 
between 100 and 240~$\mathrm{km\ s^{-1}}$. The velocity dispersions of the broad and very broad
Gaussian profiles used to model the
broad H$\alpha$ and H$\beta$ components are in the range 306-652~$\mathrm{km\ s^{-1}}$ and 1060-2100~$\mathrm{km\ s^{-1}}$, respectively. The narrowest line profiles in the two broad line components are shown by \object{TON S180}. 
Since the 
narrow-line contribution to the Balmer lines in \object{TON S180} and \object{MCG-05-01-013} is not well-defined, these two sources were fitted without
any narrow H$\alpha$, H$\beta$ or [\ion{N}{2}] components.
Including a narrow-line contribution in the 
fit can result in broader broad-line components, as these would be more confined to the broad base of the line. Regarding the values discussed in
this section, broader broad-line components would be reflected in
larger $\mathrm{FWHM(H\beta)}$, larger BH mass, and smaller Eddington ratios.

The [\ion{S}{2}]$\lambda 6716, 6731$ 
lines in \object{MS~22549-3712} were found to be redshifted 
by about 170~$\mathrm{km\ s^{-1}}$ with respect to the 
best-fitting central velocity of the other narrow lines and were modeled with independent kinematics. 
We investigated whether the apparent [\ion{S}{2}] redshift could rather indicate a [\ion{O}{3}] blueshift and artificially result
from our fitting method in which
all narrow components (especially those of H$\beta$ and H$\alpha$) are tied to [\ion{O}{3}]. However, we were unable to reproduce a similarly
good fit by tying the narrow components to [\ion{S}{2}] instead of [\ion{O}{3}] while fitting [\ion{O}{3}] with independent kinematics. Since the
narrow H$\beta$ and H$\alpha$ components as well as [\ion{N}{2}] cannot clearly be decomposed without further constraints from more
isolated narrow lines, an immediate measure of the relative velocities between those lines, [\ion{O}{3}], and [\ion{S}{2}] is not available.
Redshifted [\ion{S}{2}] emission could potentially be an indicator of shock activity in the unresolved NLR.

For the [\ion{O}{3}]$\lambda\lambda 4959$,5007 lines in
\object{MS~22549-3712} and \object{WPVS~007}, we included an additional blueshifted [\ion{O}{3}] component in order to take into account
the blue [\ion{O}{3}] wings. Compared to the main narrow lines, the blue [\ion{O}{3}] components are broader 
by about 120~$\mathrm{km\ s^{-1}}$ and blueshifted by 510~$\mathrm{km\ s^{-1}}$ and 
240~$\mathrm{km\ s^{-1}}$ in MS~22549-3712 and WPVS~007, respectively.

Table~\ref{table:AGNpars} shows the BH masses, Eddington ratios, and 
[\ion{O}{3}]$\lambda 5007$ as well as bolometric luminosities derived from the fits. 
FWHM(H$\beta$) and $\sigma$(H$\beta$) are based on the full model of the 
broad H$\beta$ component, i.e. including both, the broad and very broad Gaussian components reported in Table~\ref{table:fluxes}. 
The line shapes of AGN broad lines have been found to
correlate with line width, in the sense that FWHM/$\sigma$ increases with increasing FWHM of the line \citep{Peterson:2011aa, Kollatschny:2011aa}.
For the five objects studied here, the FWHM/$\sigma$ ratios from our broad H$\beta$ models range between 1.1 and 1.4. These ratios are 
clearly different from a Gaussian profile with 
FWHM/$\sigma$$\sim$2.35, but agree with the results for other narrow-line AGN in the FWHM(H$\beta) \lesssim 2000\ \mathrm{km\ s^{-1}}$ regime
\citep[e.g. Fig. 7 in][]{Kollatschny:2013aa}. The origin of the broad-line profiles and the best line-width indicator for virial BH mass measurements have
been a long-standing controversy which is beyond the scope of this paper. 
As in \citet{Rakshit:2017aa}, we computed the BH masses based on
FWHM(H$\beta$) together with the relation
\begin{eqnarray}
\begin{array}{l}
\log M_\mathrm{BH}= \\
\log \left[\left(\frac{\mathrm{FWHM(H\beta)}}{1000\ \mathrm{km\ s^{-1}}}\right)^2 \left(\frac{\lambda L_\lambda(5100\ \mathrm{\AA})}{10^{44}\ \mathrm{erg\ s^{-1}}}\right)^{0.533}\right]+6.69, 
\end{array}
\end{eqnarray}
where $M_\mathrm{BH}$ is the BH mass in $M_\odot$, $\mathrm{FWHM(H\beta)}$ is the full width at half maximum of 
H$\beta$ in $\mathrm{km\ s^{-1}}$, and $\lambda L_\lambda(5100\ \mathrm{\AA})$ is the 5100 \AA\ luminosity in $\mathrm{erg\ s^{-1}}$ as derived from the continuum fit to the AGN spectra (see Section~\ref{sec:AGNfit}). This relation is based on $M_\mathrm{BH}=fR_\mathrm{BLR}\Delta v^2/G$, using $f=3/4$, 
$\mathrm{FWHM(H\beta)}$ for the velocity width $\Delta v$, the gravitational constant $G$, and the best-fit radius-luminosity relation from \citet{Bentz:2013aa} in order to calculate the broad-line region radius $R_\mathrm{BLR}$ from the 5100~\AA\ luminosity.
The bolometric luminosities and Eddington ratios are estimated using 
 $L_\mathrm{bol}=9\lambda L_\lambda(5100\ \mathrm{\AA})$ \citep[e.g.][]{Kaspi:2000aa} and
 $\lambda_\mathrm{Edd} = L_\mathrm{bol}/L_\mathrm{Edd}$, respectively, where
$L_\mathrm{Edd} = 1.25\times 10^{38} M_\mathrm{BH}\ [\mathrm{erg\ s^{-1}}]$. 

The parameters of the AGN emission lines of all five objects agree with the classical criteria used to define NLS1s from optical spectra.
All objects have FWHM(H$\beta$)$< 2000\ \mathrm{km\ s^{-1}}$ -- based on the superposition of the two Gaussians used to model the broad H$\beta$ and H$\alpha$ profiles -- and
[\ion{O}{3}]/H$\beta_\mathrm{total} < 3$ (see Table~\ref{table:fluxes}).

Prominent \ion{Fe}{2} emission is typically considered as an additional characteristic of NLS1s and
all five objects show evidence of \ion{Fe}{2} emission in their nuclear spectra. 
For comparison with the large samples of NLS1s from the SDSS analyzed by 
\citet{Zhou:2006aa} and \citet{Rakshit:2017aa}, we define the \ion{Fe}{2} strength likewise as the ratio between the \ion{Fe}{2} flux,
 integrated over the wavelength range 4434-4684 \AA, and the flux of the broad H$\beta$ component. 
 As our decomposition of the nuclear spectra
 includes a ``broad'' and
 ``very broad''  Gaussian profile, we use the sum of these to compute
 $R_{4570}=$\ion{Fe}{2}$_{4434-4684\mathrm{\AA}}/\mathrm{H}\beta_\mathrm{b+vb}$ as shown in Table~\ref{table:fluxes}.
Compared to \citet{Zhou:2006aa} and \citet{Rakshit:2017aa}, the resulting \ion{Fe}{2} (4434-4684\ \AA) strength in the five objects agrees with the typical values found for NLS1s.
For their NLS1 sample, \citet{Zhou:2006aa} find an average \ion{Fe}{2} strength of $R_{4570}=0.83$, which they report to be about twice as high as
 the $R_{4570}\sim 0.4$ of normal AGN. By comparing the NLS1 sample with the BLS1s from their SDSS parent sample, \citet{Rakshit:2017aa}
 find a mean \ion{Fe}{2} strength of 0.64 for the NLS1s compared to 0.38 for the BLS1s.  
As can be seen in Table~\ref{table:fluxes}, the five NLS1s studied here show a range of $R_{4570}$ values between $R_{4570}=0.46$ 
for \object{MCG-05-01-013} and $R_{4570}=0.89$ for \object{TON S180}.

In addition to prominent \ion{Fe}{2} emission, NLS1s have been suggested to show, on average, smaller BH masses and higher Eddington ratios 
than their broad-line counterparts \citep[e.g. ][ and references therein]{Komossa:2008aa, Rakshit:2017aa}. The five objects generally 
probe the NLS1 parameter space in terms of Eddington ratios and BH masses. 
As listed in Table~\ref{table:AGNpars}, the BH masses of the objects in our study range from  
$M_\mathrm{BH} = 10^{6.43}\ M_\odot$ for \object{WPVS 007} to $M_\mathrm{BH} = 10^{6.88}\ M_\odot$ for \object{TON S180}.
The NLS1s and BLS1s samples drawn from the SDSS Data Release 12 by \citet{Rakshit:2017aa} result in  
mean BH masses of $M_\mathrm{BH}=10^{6.9}\ M_\odot$ and $10^{8}\ M_\odot$ for NLS1s and BLS1s, respectively. The corresponding dispersions of the $\log(M_\mathrm{BH})$
distributions are $0.41\ M_\odot$ for the NLS1s and $0.46\ M_\odot$ for the BLS1s. Compared to these samples, the BH masses of 
the five objects agree with the typical BH mass range displayed by NLS1s and are lower than the BH masses typically found for BLS1s.

The highest Eddington ratios among the five sources are displayed by 
\object{TON S180} ($\lambda_\mathrm{Edd} =1.7$ or $\log \lambda_\mathrm{Edd} = 0.2$) and \object{WPVS 007} ($\lambda_\mathrm{Edd} =0.58$ or $\log \lambda_\mathrm{Edd} = -0.2$), while  
\object{ESO 399-IG20}, \object{MCG-05-01-013}, and \object{MS 22549-3712} show Eddington ratios below 0.24 ($\log \lambda_\mathrm{Edd} < -0.62$). 
Compared to the NLS1s studied by \citet{Rakshit:2017aa}, the Eddington ratios of \object{TON S180} and \object{WPVS 007} overlap with the upper and average portion of the NLS1 Eddington ratio distribution, respectively. The Eddington ratios of the other three sources agree with the lower part of the distribution.

An estimate of the electron density in the NLR based on the density-sensitive [\ion{S}{2}] is available for three sources
for which the [\ion{S}{2}] lines could be fitted in the nuclear spectra. Assuming an electron temperature of
$T_e=10,000~K$, \object{ESO~399-IG20} and \object{WPVS 007} show NLR densities of about $n_e\approx120\ \mathrm{cm^{-3}}$ and
 $n_e\approx240\ \mathrm{cm^{-3}}$, respectively. The [\ion{S}{2}]$\lambda$6716/[\ion{S}{2}]$\lambda$6731 ratio measured for
 \object{MS 22549-3712} falls into the low-density limit of $\sim 10\ \mathrm{cm^{-3}}$. 
 According to \citet{Xu:2007aa}, NLS1s can have very
 low densities in their NLRs, in contrast to BLS1s which typically show $n_e \gtrsim 140\ \mathrm{cm^{-3}}$. The samples of NLS1s and BLS1s studied by 
 \citet{Rakshit:2017aa} do not confirm such a lower-density limit for BLS1s, but suggest a weak trend of lower-density NLRs in NLS1s compared to BLS1s.
 The electron density for \object{WPVS 007} overlaps with the bulk of NLS1 and BLS1 electron densities from that study, while the NLR densities for
 \object{ESO~399-IG20} and \object{MS 22549-3712} agree with the lower-density wing of the distribution presented by \citet{Rakshit:2017aa}.

\floattable
\begin{deluxetable}{lc|ccccc}
\tablecaption{Line fluxes and kinematics determined from the host-deblended AGN spectra \label{table:fluxes} }
\tablecolumns{8}
\tablewidth{0pt}
\tablehead{
\colhead{Line} & \colhead{Wavelength} & \colhead{\object{ESO~399-IG20}}    & \colhead{\object{MCG-05-01-013}} & \colhead{\object{MS 22549-3712}}  & \colhead{\object{TON S180}} & \colhead{\object{WPVS 007}}  \\
 \colhead{}       &  \colhead{(\AA)}     &\colhead{}     }
\startdata
        &              & \multicolumn{5}{c}{Flux ($\mathrm{10^{-16}\ erg\ s^{-1}\ cm^{-2}}$) }     \\                
\ion{Fe}{2}                             &$ 4434-4684$&$ 1060\pm 20  $&$180\pm 10  $&$   410\pm10   $&$ 1670\pm 70    $&$ 820\pm 10 $\\      
\ion{He}{2}                             &$  4685.7     $&$  143\pm 9   $&$ 13\pm 5       $&$   89\pm 7       $&$ 120\pm 30      $&$ 17\pm 6$\\
H$\beta_\mathrm{n}$            &$ 4861.3       $&$  49\pm 6      $&$ -                 $&$  49\pm 7       $&$ -     $&$ 20\pm 4$\\
H$\beta_\mathrm{b}$            &$ 4861.3       $&$  810\pm 20  $&$ 190\pm 9   $&$  510\pm 20   $&$ 530\pm 50   $&$ 750\pm 10$\\
H$\beta_\mathrm{vb}$          &$ 4861.3       $&$  1090\pm 30 $&$ 200\pm 10 $&$  160\pm 20  $&$ 1340\pm 80   $&$ 600\pm 20$\\
$[$\ion{O}{3}$]_\mathrm{bl}$  &$ 4958.8     $&$ -                    $&$ -                 $&$  50\pm 20    $&$  -                   $&$ 50\pm 10$\\
$[$\ion{O}{3}$]$                     &$ 4958.8      $&$ 178\pm 6       $&$ 15\pm 3     $&$  70\pm 20    $&$ 120\pm 30    $&$ 30\pm 10$\\
$[$\ion{O}{3}$]_\mathrm{bl}$  &$ 5006.8     $&$ -                    $&$ -                 $&$  150\pm 20  $&$  -                   $&$ 140\pm 10$\\
$[$\ion{O}{3}$]$                     &$ 5006.8       $&$ 531\pm 6      $&$ 46\pm 3     $&$  200\pm 20  $&$ 360\pm 30   $&$ 90\pm 10$\\
$[$\ion{O}{1}$]$                        &$ 6300.2   $&$ 27\pm 5        $&$ -                 $&$  10\pm 4     $&$ 30\pm 20    $&$ 9\pm 3$\\
$[$\ion{O}{1}$]$                        &$ 6363.7   $&$ 30\pm 5        $&$ -                 $&$  26\pm 4     $&$ 30\pm 20    $&$ 15\pm 3$\\
$[$\ion{Fe}{10}$]$                     &$ 6374.5   $&$ 37\pm 5        $&$ -                 $&$  22\pm 4     $&$  -                  $&$ 2\pm 3$\\
$[$\ion{N}{2}$]$                         &$ 6548.0   $&$ 32\pm 7        $&$ -                 $&$  67\pm 10   $&$  -                  $&$ 31\pm 4$\\
H$\alpha_\mathrm{n}$             &$ 6562.8    $&$ 82\pm 7        $&$ -                 $&$  270\pm 20 $&$  - $&$ 301\pm 8$\\
H$\alpha_\mathrm{b}$             &$ 6562.8    $&$ 3840\pm 60  $&$ 1100\pm 30$&$ 1770\pm 50$&$ 1950\pm 70 $&$ 2650\pm 20$\\
H$\alpha_\mathrm{vb}$           &$ 6562.8    $&$ 2710\pm 50  $&$ 740\pm 30  $&$  610\pm 30 $&$ 3420\pm 70 $&$ 1640\pm 30$\\
$[$\ion{N}{2}$]$                        &$ 6583.3    $&$ 96\pm 7        $&$ -                 $&$  200\pm 10 $&$  -                 $&$ 94\pm 4$\\
\ion{He}{1}                                &$ 6678.2    $&$  120\pm 10   $&$ -                 $&$  41\pm 7   $&$ 40\pm 20       $&$ 48\pm 6$\\
$[$\ion{S}{2}$]$                        &$ 6716.3    $&$ 80\pm 5        $&$ -                 $&$  33\pm 3    $&$ -                   $&$ 47\pm 3$\\
$[$\ion{S}{2}$]$                        &$ 6730.7    $&$ 59\pm 5        $&$ -                 $&$  20\pm 3    $&$ -                   $&$ 39\pm 3$\\
\multicolumn{2}{l|}{$[$\ion{O}{3}$]\lambda5007/\mathrm{H}\beta_\mathrm{total}$} & $0.272\pm 0.006$&$0.118\pm0.009$&$ 0.49\pm 0.04 $&$ 0.19\pm0.02 $&$ 0.17\pm 0.01$\\
\multicolumn{2}{l|}{$R_{4570}=$\ion{Fe}{2}$_{4434-4684\mathrm{\AA}}/\mathrm{H}\beta_\mathrm{b+vb}$} &$ 0.56\pm 0.01 $& $ 0.46\pm 0.03 $ & $0.61\pm 0.03 $&$0.89\pm 0.06$&$0.61\pm 0.01$ \\
\multicolumn{2}{l|}{$[$\ion{S}{2}$]\lambda6716/[$\ion{S}{2}$]\lambda6731$} & $1.3\pm 0.1$&$ - $&$ 1.7\pm 0.3 $&$ - $&$ 1.2\pm 0.1$\\
\hline                                   
                                            &                &         \multicolumn{5}{c}{Systemic velocity ($\mathrm{km\ s^{-1}}$)}       \\[0.5ex]
\multicolumn{2}{l|}{Narrow component\tablenotemark{a}}      &$7545\pm 2    $&$ 9161\pm 9   $&$11664\pm 4    $&$18430\pm 20   $&$ 8585\pm 1 $ \\
                                             &                &         \multicolumn{5}{c}{Relative velocity ($\mathrm{km\ s^{-1}}$)}       \\[0.5ex]
\multicolumn{2}{l|}{Broad component\tablenotemark{b}}         &$+208\pm 4    $&$+150\pm  10    $&$+250\pm 7          $&$+200\pm 20      $&$ +83\pm 2 $\\
\multicolumn{2}{l|}{Very broad component\tablenotemark{c}} &$+200\pm 20   $&$+360\pm  50  $&$+10\pm 70            $&$+240\pm 30 $&$ -140\pm 20  $\\
\multicolumn{2}{l|}{Blueshifted [\ion{O}{3}]\tablenotemark{d}} &$   -               $&$      -              $&$-510\pm 50         $&$     -                $&$ -240\pm 20 $\\
\multicolumn{2}{l|}{[\ion{S}{2}]\tablenotemark{f}}                     &$   -               $&$      -              $&$+170\pm 10           $&$     -                $&$- $\\
\hline                                   
                                             &                &         \multicolumn{5}{c}{Velocity dispersion ($\mathrm{km\ s^{-1}}$)}       \\[0.5ex]
\multicolumn{2}{l|}{Narrow component\tablenotemark{a}}      &$148\pm 2      $&$130\pm 10      $&$231\pm 6       $&$220\pm 20      $&$ 98\pm 2 $ \\
\multicolumn{2}{l|}{Broad component\tablenotemark{b}}         &$636\pm 6     $&$652\pm 9        $&$645\pm 8       $&$306\pm 7   $&$ 412\pm 3 $\\
\multicolumn{2}{l|}{Very broad component\tablenotemark{c}} &$1830\pm 30  $&$ 2070\pm 70  $&$2100\pm 100 $&$1060\pm 20 $&$ 1390\pm 20  $\\
\multicolumn{2}{l|}{Blueshifted [\ion{O}{3}]\tablenotemark{d}} &$   -                $&$  -            $&$350\pm 40     $&$     -                $&$ 220\pm 10 $\\
\multicolumn{2}{l|}{[\ion{S}{2}]\tablenotemark{e}}                     &$   -                $&$ -              $&$99\pm 9        $&$     -                $&$ - $\\
\enddata
\tablenotetext{a}{The narrow component is used as reference for the systemic velocity. The narrow component includes H$\beta_\mathrm{n}$, [\ion{O}{3}], [\ion{N}{1}], [\ion{O}{1}], [\ion{Fe}{10}], [\ion{N}{2}], 
H$\alpha_\mathrm{n}$, and [\ion{S}{2}], except in the case of MS 22549-3712 where the [\ion{S}{2}] lines were fitted with independent kinematics.} 
\tablenotetext{b}{Broad components include H$\beta_\mathrm{b}$, H$\alpha_\mathrm{b}$, \ion{He}{2}, and \ion{He}{1} .} 
\tablenotetext{c}{Very broad components include H$\beta_\mathrm{vb}$ and H$\alpha_\mathrm{vb}$.}
\tablenotetext{d}{Blueshifted [\ion{O}{3}] component corresponds to [\ion{O}{3}]$_\mathrm{bl}$.}
\tablenotetext{e}{The [\ion{S}{2}] lines were fitted with independent kinematics in the case of MS 22549-3712, while in all other cases they share the kinematics of the ``narrow component''.}
\end{deluxetable}

\floattable
\begin{deluxetable}{lcccccc}
\tablecaption{Parameters derived from the host-deblended AGN spectra \label{table:AGNpars}  }
\tablecolumns{7}
\tablewidth{0pt}
\tablehead{ 
\colhead{Object} & \colhead{FWHM(H$\beta$)} & 
\colhead{$\sigma(\mathrm{H\beta})$} & \colhead{$\log(M_\mathrm{BH})$} & 
\colhead{$\log(L([$\ion{O}{3}$])$} & \colhead{$\log(L_\mathrm{bol})$} & 
\colhead{$\lambda_\mathrm{Edd}$}  \\
\colhead{}          &  \colhead{$\mathrm{(km\ s^{-1})}$} &  
\colhead{$\mathrm{(km\ s^{-1})}$} & \colhead{$(M_\odot)$ }  &   
\colhead{$\mathrm{(erg\ s^{-1})}$}   &   \colhead{$\mathrm{(erg\ s^{-1})}$}  & \colhead{} }
\decimalcolnumbers
\startdata
 \object{ESO~399-IG20}    &$ 1900\pm 20 $&$ 1450\pm 20$&$ 6.77\pm 0.02 $&$40.854\pm 0.005$&$ 44.05\pm 0.04$&$  0.15\pm 0.02 $\\  
  \object{MCG-05-01-013} &$ 1850\pm 40 $&$ 1560\pm 60$&$ 6.53\pm 0.04 $&$39.97\pm 0.02     $&$ 43.65\pm0.08$&$  0.11\pm 0.02 $\\
  \object{MS 22549-3712}  &$ 1610\pm 20 $&$ 1190\pm 60$&$ 6.67\pm0.02  $&$41.08\pm 0.04      $&$44.14\pm 0.05$&$ 0.24\pm 0.03  $\\
 \object{TON S180}           &$ 1060\pm50 $&$  920\pm 20 $&$ 6.88\pm 0.05   $&$ 41.51\pm 0.04     $&$45.21\pm 0.04$&$ 1.7\pm 0.2 $\\
    \object{WPVS 007}         &$ 1120\pm10 $&$   980\pm 10$&$ 6.43\pm 0.01  $&$40.62\pm 0.03    $&$44.29\pm0.02$&$ 0.58\pm 0.03  $\\
 \enddata
\tablecomments{Column 1: Object name; Columns 2 and 3: FWHM and second moment of the 
broad H$\beta$ line taking into account the superposition of the broad and very broad Gaussian components listed in Table~\ref{table:fluxes};
Column 4: Logarithm of the derived BH mass; Column 5: Logarithm of the total
[\ion{O}{3}]$\lambda 5007$ luminosity (including the blueshifted component, where applicable); Column 6: Logarithm of the bolometric luminosity, based on the 5100\ \AA\ flux; Column 7: Eddington ratio.
A detailed description of how the quantities are derived is given in the text. The errors noted in the table are the formal errors 
propagated from the line fitting process. The actual uncertainties in the flux-dependent quantities are likely to be larger, since we conservatively estimate a factor
of 2 of uncertainty in the flux calibration (see Section~\ref{sec:obs}). This corresponds to uncertainties of $\log(\sqrt{2})\approx 0.2~\mathrm{dex}$ for $\log(M_\mathrm{BH})$,
$\log(2)\approx 0.3~\mathrm{dex}$ for $\log(L([$\ion{O}{3}$]))$ and $\log(L_\mathrm{bol})$, and a factor of $\sqrt{2}\approx 1.4$ for 
$\lambda_\mathrm{Edd}$.}
\end{deluxetable}

\subsection{Properties of the host galaxies}\label{sec:host}

\begin{figure*}
\figurenum{3}
\gridline{\fig{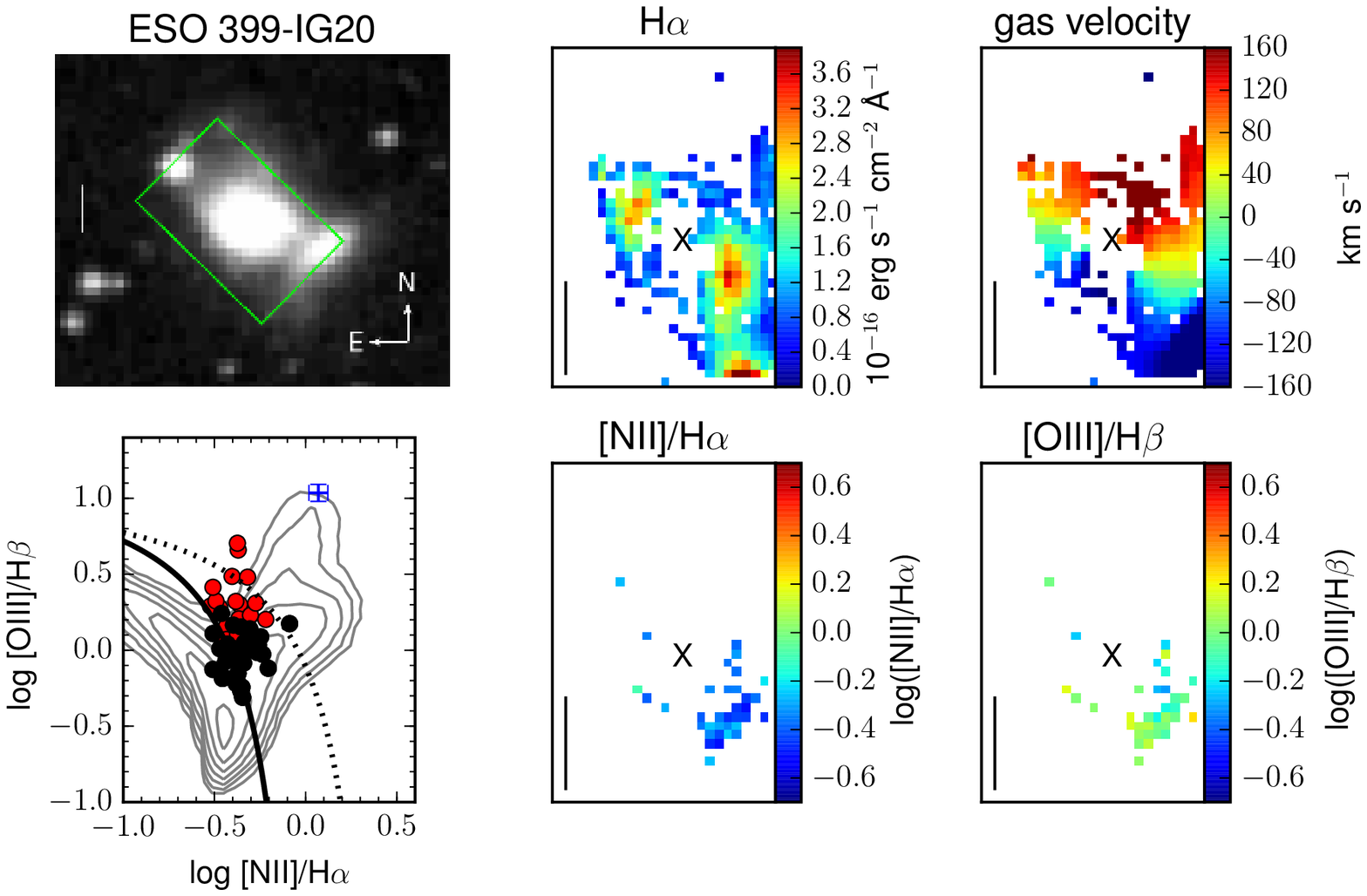}{\textwidth}{(a) ESO~399-IG20. The WiFeS maps show north-east up and north-west to the right. 
The companion galaxy, visible at the western corner of the H$\alpha$ and velocity maps, has been masked in the line ratio maps (lower middle and right panels) and is marked with red data points in the diagnostic diagram
(lower left panel).}}
\gridline{\fig{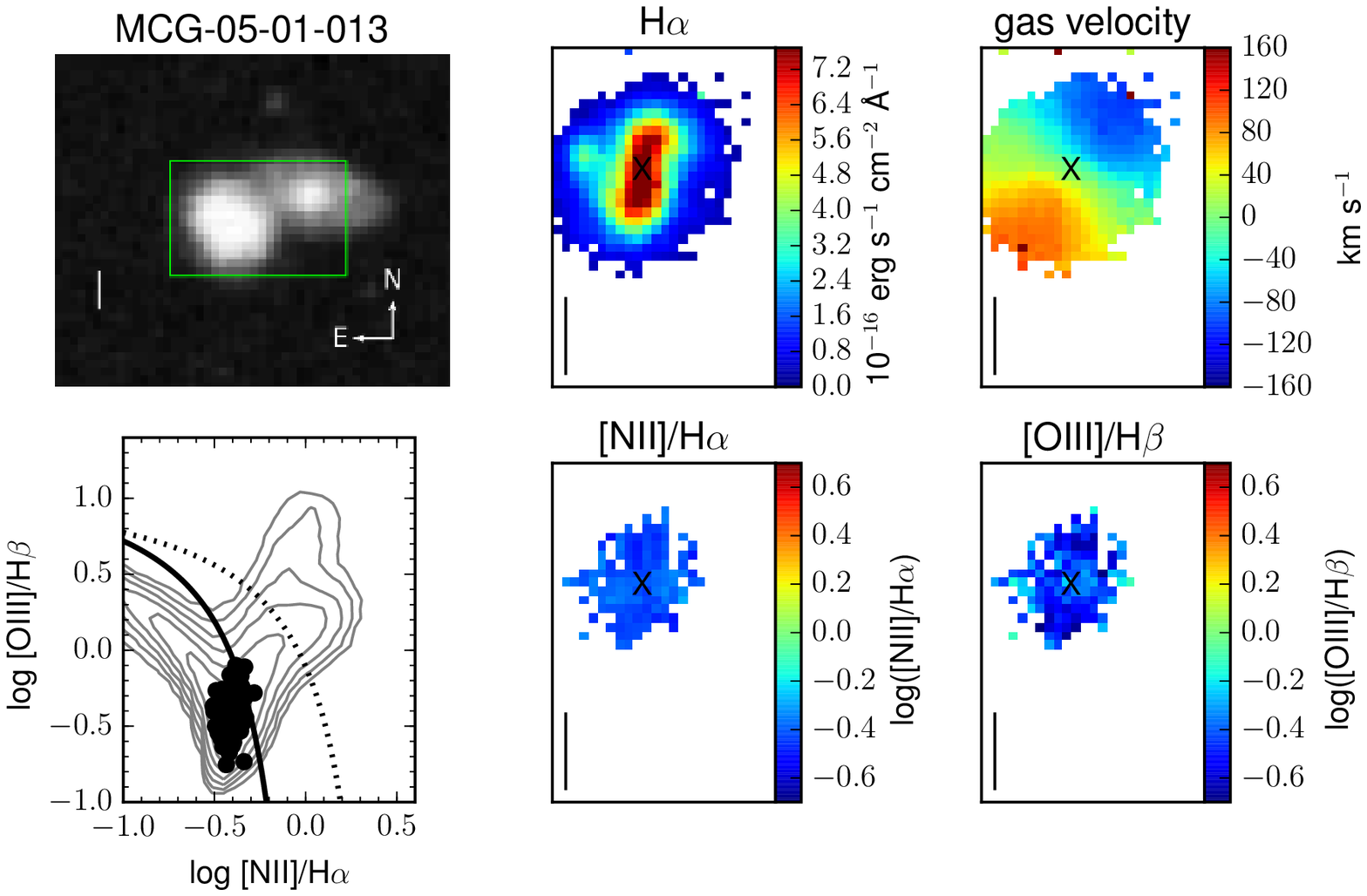}{\textwidth}{(b) MCG-05-01-013. The WiFeS maps show east up and north to the right. The AGN line ratio is not included in the lower left panel, because the [\ion{N}{2}] and narrow Balmer lines could not be clearly detected in the nuclear
AGN spectrum (see Table~\ref{table:fluxes} and Fig.~\ref{fig:nuclfit}b).}}
\end{figure*}  
\begin{figure*}
\figurenum{3}
\gridline{\fig{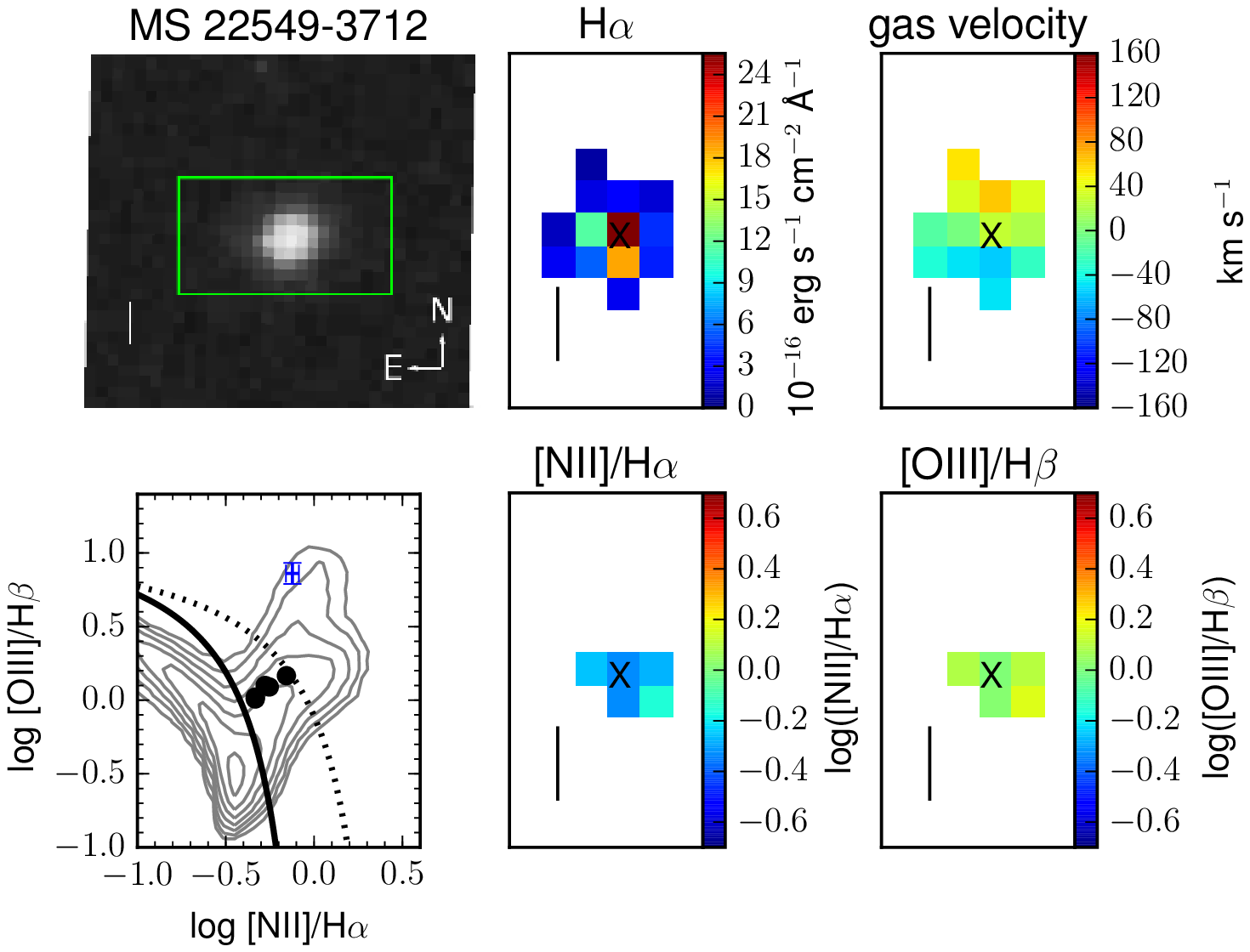}{\textwidth}{(c) MS 22549-3712. The WiFeS maps show east up and north to the right.}
         }
\gridline{\fig{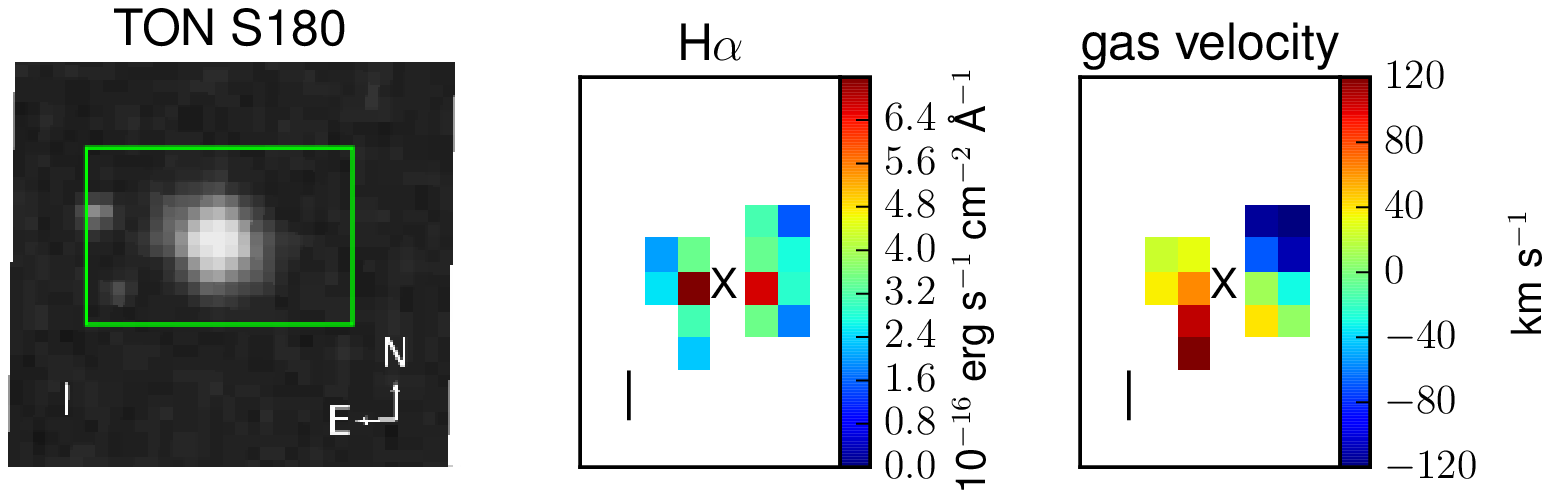}{\textwidth}{(d) TON S180. The WiFeS maps show east up and north to the right. Based on the
available data set, a line ratio analysis
for this source was not possible.}
         }
\end{figure*}  
\begin{figure*}
\figurenum{3}
\gridline{\fig{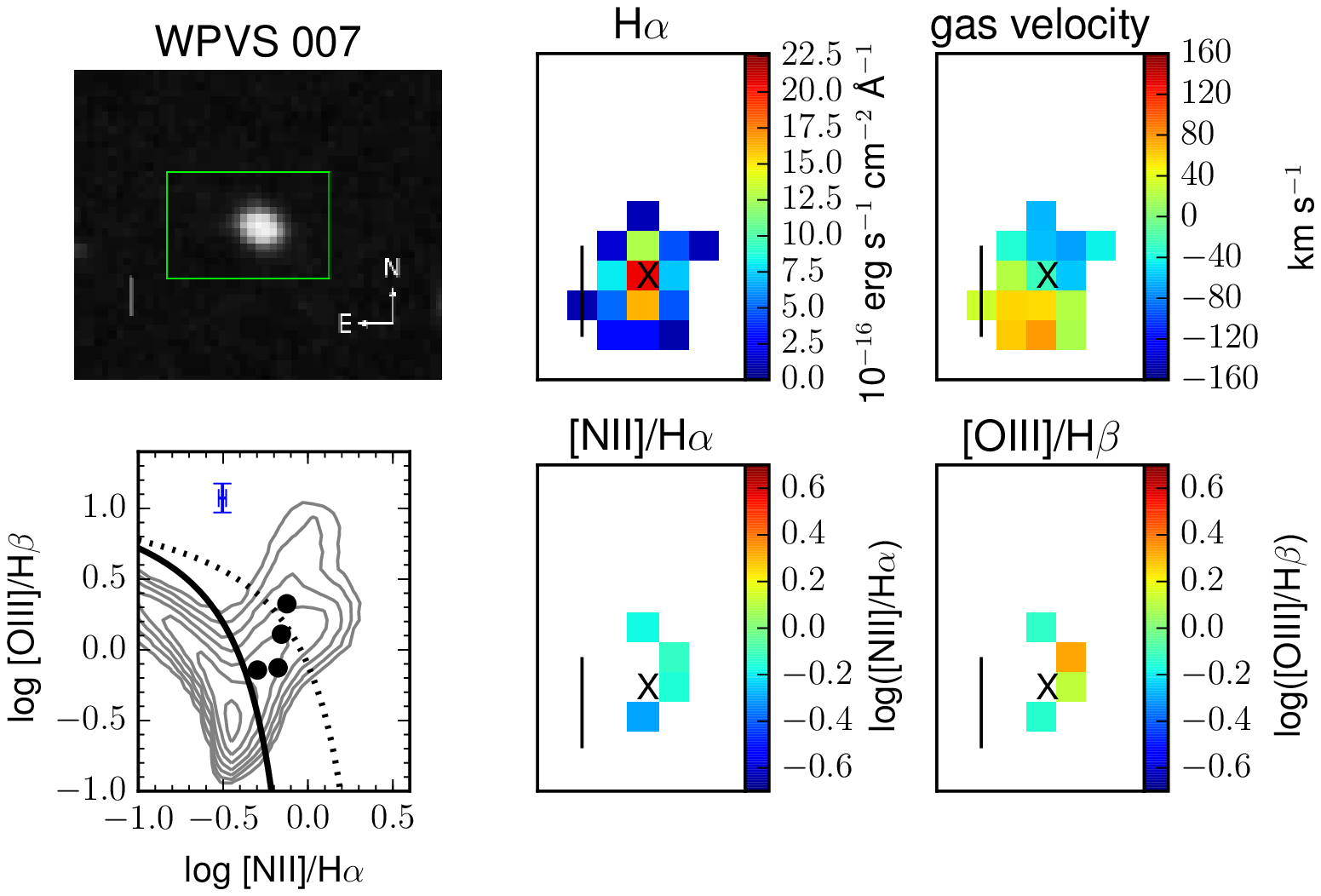}{\textwidth}{(e) WPVS 007. The WiFeS maps show east up and north to the right.}
         }
 \caption{Flux, kinematic, and diagnostic maps for the NLS1 host galaxies. All maps are derived from the host galaxy data cubes after
  AGN-host galaxy separation. Top left panel: DSS image with an overlay of the WiFeS field-of-view; Top middle panel: H$\alpha$ flux map;
  Top right panel: Gas velocity map; Bottom left panel: Line ratio diagram in [\ion{O}{3}]$\lambda 5007$/H$\beta$ and [\ion{N}{2}]$\lambda 6583$/H$\alpha$ 
  for all host-galaxy pixels with sufficient signal-to-noise ratio (black dots) and for the narrow-line component from the nuclear AGN spectrum (blue error bar), compared to
  SDSS galaxies from the SDSS MPA/JHU galaxy catalog \citep{Brinchmann:2004aa}, (grey contours); Bottom middle panel: Line ratio map in 
  [\ion{N}{2}]$\lambda 6583$/H$\alpha$; Bottom right panel: Line ratio map in [\ion{O}{3}]$\lambda 5007$/H$\beta$. 
    The black solid and black dotted lines in the line ratio diagrams (bottom left panels) represent the empirical and theoretical maximum starburst lines
    from \citet{Kauffmann:2003aa} and \citet{Kewley:2001aa}, respectively.
    The scale bars in the maps indicate 5~kpc. The black cross indicates the position of the AGN, as determined from an elliptical Gaussian fit to the host-cleaned AGN cube
    integrated over the [\ion{O}{3}]$\lambda 5007$ line. Data points with a signal-to-noise ratio $<3$, an absolute velocity error of $>30\ \mathrm{km\ s^{-1}}$, and an absolute velocity dispersion error of $>80\ \mathrm{km\ s^{-1}}$ 
    ($>200\ \mathrm{km\ s^{-1}}$ for \object{TON S180}) have been clipped.
 The data cubes for \object{MS 22549-3712},
  \object{TON S180}, and \object{WPVS 007} were binned to $3\arcsec \times 3\arcsec$ pixels in order to increase the signal-to-noise ratio.
    \label{fig:maps} }
\end{figure*}

The emission-line maps derived from the AGN-deblended data cubes for the host galaxies of the five objects are shown in Fig.~\ref{fig:maps}. 
All resulting gas velocity fields show evidence of large scale gas rotation. Except for \object{MCG-05-01-013}, which is obviously a disk galaxy with a bar
and/or spiral arm, the spatial resolution is too low in order to unambiguously determine the galaxy type.
However, \object{MS 22549-3712} and \object{TON S180} are known to be disk galaxies from high-resolution HST images which suggest a
pseudo-bulge in \object{MS 22549-3712} and a classical bulge in \object{TON S180} \citep{Mathur:2012aa}. \object{WPVS 007} was found to be 
a largely bulge-dominated galaxy with possible evidence of sub-structure and a low S\'{e}rsic index \citep{Busch:2014aa}.
\object{ESO 399-IG20} was classified as S0 galaxy by \citet{Dietrich:2005aa} based on spectral fitting. 
The gas velocity field of \object{ESO 399-IG20} is more perturbed, possibly as a consequence of a tidal interaction with the companion to the south-west. The companion, which is evident from the DSS image, is associated with the H$\alpha$ peak in the south-western-most
corner of the WiFeS H$\alpha$ map.
The $g-r$ colors of the five host galaxies, presented in Section~\ref{sec:sfr}, are generally consistent with these galaxy-type classifications 
\citep[cf. e.g.][]{Fernandez-Lorenzo:2012aa}.
\citet{Orban-de-Xivry:2011aa} proposed that the nuclear activity in NLS1s might be driven by secular evolution. Based on the gas kinematics and  
suggested galaxy types, it is possible that secular
evolution plays a role in at least some of the five objects studied here.
The data for \object{MS 22549-3712},
\object{TON S180}, and \object{WPVS 007} have been binned spatially by $3\times3$ pixels ($3\arcsec \times3\arcsec$) in order to increase the signal-to-noise
ratio on the emission lines. Despite the lower spatial resolution in these data, the gas kinematics is likewise characterized by velocity gradients centered on the
galaxy nuclei, presumably indicating rotation.

According to the line ratio diagnostics presented in the bottom rows of Fig.~\ref{fig:maps}, none of the five objects shows
clear evidence of an extended NLR at the 2-3 kpc scales probed by our observations.
The line ratios for the five sources largely occupy the region dominated by 
ionization from \ion{H}{2} regions close to or below the empirical maximum starburst line from \citet{Kauffmann:2003aa}, (solid black line). The binned data for \object{MS 22549-3712} and \object{WPVS 007} show marginal 
evidence of off-nuclear spaxels extending into the region around and above the theoretical maximum starburst line from \citet{Kewley:2001aa},
which could indicate some AGN or shock contribution to the gas ionization 
\citep[e.g.][]{Rich:2011aa, Scharwachter:2011ab, Davies:2014aa, Dopita:2014aa}. 
A possible shock activity in \object{MS 22549-3712} has also been pointed out based on the redshifted [\ion{S}{2}] lines in the nuclear spectrum of this
object (Section~\ref{sec:obs:nuclfit}). 
However, this interpretation remains speculative
given the low signal-to-noise ratio in the host galaxy data for these objects. 

\newpage

\subsection{Star formation rates and stellar mass}\label{sec:sfr}

\floattable
\begin{deluxetable}{lcccccc}
\tablecaption{Star formation rate and host galaxy stellar mass \label{table:sfr}  }
\tablecolumns{7}
\tablewidth{0pt}
\tablehead{ 
\colhead{Object} & \colhead{$\log L_\mathrm{H\alpha}$} & 
\colhead{SFR } & \colhead{$g-r$} & 
\colhead{$m_V$} & \colhead{$\log M_\star$} & 
\colhead{$\log\mathrm{sSFR}$}  \\
\colhead{}          &  \colhead{$(\mathrm{erg\ s^{-1}})$} &  
\colhead{$\mathrm{(M_\odot\ yr^{-1})}$} & \colhead{(mag) }  &   
\colhead{(mag)}   &   \colhead{$(M_\odot)$}  & \colhead{$(\mathrm{yr^{-1}})$} }
\decimalcolnumbers
\startdata
 \object{ESO~399-IG20}    &  $40.84\pm 0.04$$^\mathrm{a}$  & $0.55\pm 0.05$ &  0.59 & $14.1\pm0.8$ & $10.8\pm 0.3$   & $-11.0\pm 0.3$ \\     
  \object{MCG-05-01-013}  & $41.55\pm 0.01$$^\tablenotemark{a}$  & $2.82\pm 0.08$        & 0.44  & $14.6\pm0.8$ & $10.6\pm 0.3$  & $-10.2\pm 0.3$ \\
  \object{MS 22549-3712}   & $40.8\pm 0.4$$^\tablenotemark{b}$  & $0.5\pm 0.4$   & 0.75  & $16.2\pm 0.8$ & $10.5\pm 0.3$ & $-10.8\pm 0.4$ \\
  \object{TON S180}            & $41.0\pm 0.4$$^\tablenotemark{b}$  & $0.8\pm 0.6\tablenotemark{c}$   & 0.40  & $15.7\pm0.8$ & $10.7\pm 0.3$ & $-10.8\pm 0.4$\tablenotemark{c} \\
    \object{WPVS 007}          & $40.6\pm 0.4$$^\tablenotemark{b}$  & $0.3\pm 0.2$  & 0.40 & $16.5\pm0.8$ &  $9.6\pm 0.3$    & $-10.1\pm 0.4$ \\
 \enddata
\tablecomments{Column 1: Object name; Column 2: Extinction-corrected H$\alpha$ luminosity; 
Column 3: H$\alpha$-based star formation rate (SFR); 
Column 4: $g-r$ color of the host galaxy; 
Column 5: $V$-band magnitude of the host galaxy; Column 6: Stellar mass of the host galaxy;
Column 7: Logarithm of the specific star formation rate. 
For $\log L_\mathrm{H\alpha}$ and the SFR, we report the formal errors propagated from the line fitting. 
For the stellar mass, we assume an error
of 0.3~dex, corresponding to the estimated factor
of 2 of uncertainty in the flux calibration (see Section~\ref{sec:obs}). Note that the same uncertainty due to the flux calibration may also apply to the
H$\alpha$ luminosity and star formation rate.}
\tablenotetext{a}{Integrated H$\alpha$ luminosity based on a pixel-by-pixel extinction correction, using all pixels with reliable $A_V$ as shown in Fig.~\ref{fig:av}.}
\tablenotetext{b}{Integrated H$\alpha$ luminosity based on a global extinction correction assuming an average $A_V=1\pm 1$.}
\tablenotetext{c}{Upper limit (assuming that the H$\alpha$ luminosity in the host galaxy of \object{TON S180} is indicative of star formation)}
\end{deluxetable}    
    
 \begin{figure}
\figurenum{4}
\plotone{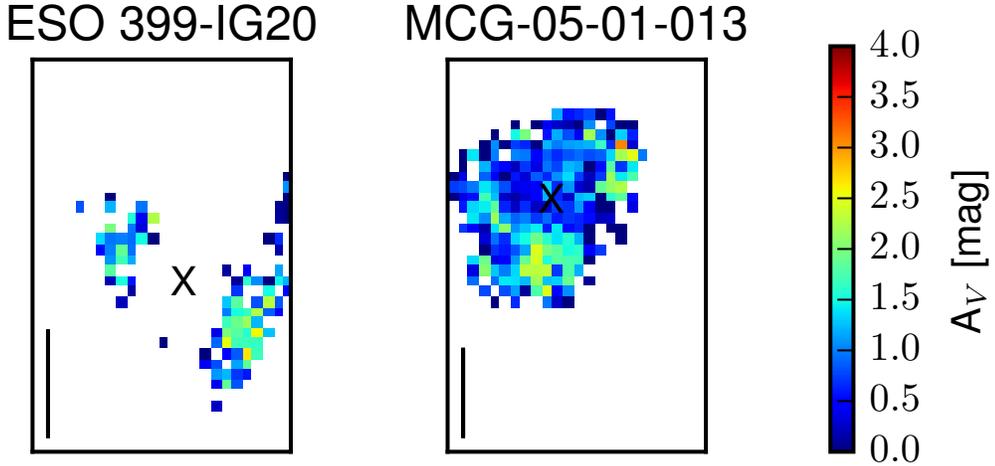}
\caption{Extinction maps for the two objects with well-resolved data. The scale bars in the maps indicate 5~kpc. The black cross shows the position of the AGN, as determined from an elliptical Gaussian fit to the host-cleaned AGN cube
    integrated over the [\ion{O}{3}]$\lambda 5007$ line. The orientation of the maps is the same as in Fig.~\ref{fig:maps}. 
    The $A_V$ shown here is used for computing the total H$\alpha$ luminosity and
  global star formation rate based on a pixel-by-pixel extinction correction. \label{fig:av}}
\end{figure}   

Estimates for the global star formation rates, stellar masses, and specific star formation rates are listed in Table~\ref{table:sfr}.
Since the extended ionized gas in all five objects is predominantly excited by star formation, the global star formation rates are 
derived from the integrated H$\alpha$ luminosities, $L_\mathrm{H\alpha}$, using the relation from \citet{Kennicutt:1998aa}
\begin{equation}
\mathrm{SFR} \ (M_\odot\ \mathrm{yr^{-1}}) = \frac{L_\mathrm{H\alpha}}{1.26\times 10^{41}\ \mathrm{erg\ s^{-1}}}.
\end{equation}
The H$\alpha$ luminosities are corrected for extinction based on the Balmer decrement using an attenuation correction of the form
\begin{equation}
\frac{\tau_\lambda}{\tau_V}=(1-\mu)\left( \frac{\lambda}{5500\ \mathrm{\AA}} \right)^{-1.3} + \mu \left( \frac{\lambda}{5500\ \mathrm{\AA}} \right)^{-0.7},
\end{equation}
\citep{Wild:2007aa,Wild:2011aa}, with $\mu = 0.4$, where $\lambda$ is the wavelength in \AA.
Expressed in terms of the extinction $A_V$, this corresponds to a flux correction of
\begin{equation}
f_\mathrm{H\alpha,\ corr} = f_\mathrm{H\alpha} 10^{0.4 A_V \tau_\mathrm{H\alpha}/\tau_V},
\end{equation}
where $f_\mathrm{H\alpha}$ is the flux at the wavelength of H$\alpha$.
For the spatially well-resolved sources, \object{ESO~399-IG20} and \object{MCG-05-01-013}, the extinction correction is applied on a pixel-by-pixel basis.
The corresponding extinction maps of these sources are shown in Fig.~\ref{fig:av}. Like in 
Fig.~\ref{fig:maps}, only data with a flux signal-to-noise ratio $>3$, a velocity error $<30\ \mathrm{km\ s^{-1}}$, and a velocity
dispersion error  $<80\ \mathrm{km\ s^{-1}}$ are used. Furthermore, a few unreliable $A_V$ values, which are $<0$ within the uncertainties, are masked out.
Since the calculation of $A_V$ involves the fainter H$\beta$ line, the number of pixels with reliable $A_V$ data is smaller than the number of pixels with reliable H$\alpha$ fluxes.
In order to obtain the integrated H$\alpha$ luminosity, only the H$\alpha$ fluxes in spatial pixels with reliable $A_V$ measurements were corrected for extinction and integrated.
Within the uncertainties, the same H$\alpha$ luminosities are obtained when integrating over
all available spatial pixels with reliable H$\alpha$ fluxes while adopting an assumed $A_V\sim 0-1\ \mathrm{mag}$ for the fraction of pixels lacking reliable $A_V$ measurements. 
Since the signal-to-noise ratios
for \object{MS 22549-3712}, \object{TON S180}, and \object{WPVS 007} are too low to provide reliable $A_V$ measurements, the H$\alpha$ fluxes 
for these sources were corrected assuming an average $A_V=1$~mag. This value corresponds to the average $A_V$ found for the two well-resolved
sources \object{ESO~399-IG20} and \object{MCG-05-01-013}. The star formation rates for \object{TON S180} are based on the assumption
that the H$\alpha$ emission is dominated by star formation, though the nature of the dominant ionization mechanism could not be verified
using line ratio diagrams (see Fig.~\ref{fig:maps}d). Thus, the star formation for this source may be considered as an upper limit.

In addition to the star formation rates, Table~\ref{table:sfr} shows estimates of the stellar masses of the five objects. 
These were derived from the integrated $r$-band magnitude of the best-fitting surface brightness model described in 
Section~\ref{sec:deblend}. We use a mass-to-light
ratio for the $V$-band ($\Gamma_V$) based on the $g-r$ color, following the prescription of \citet{Wilkins:2013aa},
\begin{equation}
 \log \Gamma_V = 1.4 \times (g-r) -0.49.
\end{equation}
Given the low redshift of the objects any $k$-correction would be small and was ignored for this work. We also ignore a correction
for dust attenuation, since the 
amount of dust attenuation for the stellar continuum cannot be estimated from the data themselves.

 \begin{figure}
\figurenum{5}
\epsscale{0.85}
\plotone{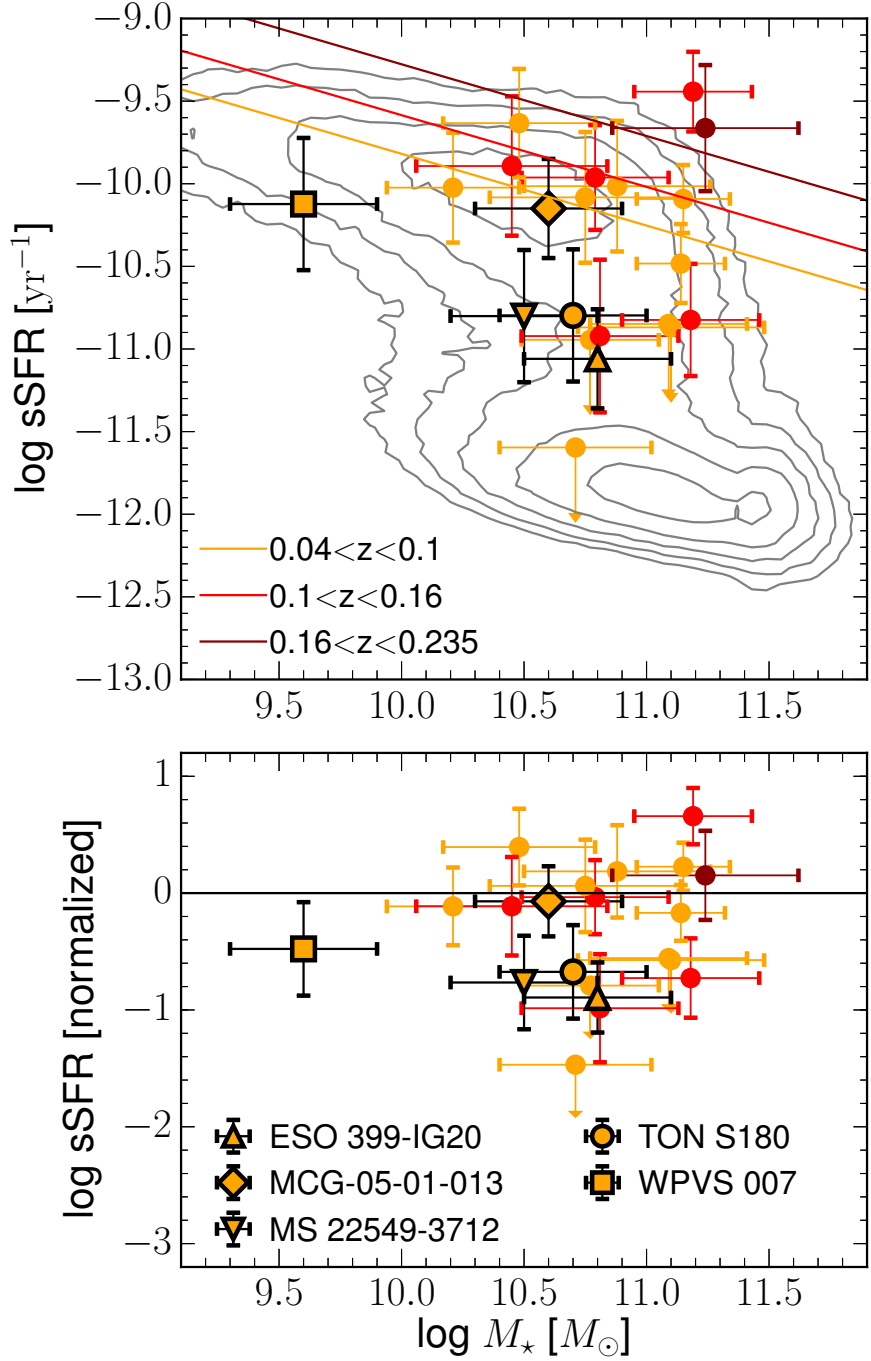}
\caption{Comparison of the specific star formation
  rates and stellar masses of the five objects with SDSS star forming galaxies, main sequence relationships from the Galaxy And Mass Assembly (GAMA) survey, and 
  a sample of $z<0.2$ type-1 QSOs. The NLS1s from this paper are shown as colored symbols with black borders and black error bars, as explained in the legend. 
  The colored circles indicate the $z<0.2$ type-1 QSOs from the sample by
   \citet{Husemann:2014aa}.
  Top panel:
  SDSS galaxies from the SDSS MPA/JHU galaxy catalogue \citep{Brinchmann:2004aa} are shown as contours. The orange, red, and dark red lines show the fits to the main sequence
  from \citet{Lara-Lopez:2013aa} for redshift bins of $0.04<z<0.1$, $0.1<z<0.16$, and $0.16<z<0.235$, respectively. The NLS1 and $z<0.2$ type-1 QSOs
  data points have been colored based 
  on their corresponding redshift bins.
  Bottom panel: This panel shows the same data points as the top panel, normalized by the main sequence fit corresponding to their redshift bin.\label{fig:sfrmass}}
\end{figure}  

In Fig.~\ref{fig:sfrmass}, the specific star formation rates and stellar masses of the five objects are 
compared to SDSS galaxies from \citet{Brinchmann:2004aa} and to the flux-limited sample of $z<0.2$
type-1 QSOs discussed by \citet{Husemann:2014aa}.
Since the comparison of the different sources 
is subject to a mild redshift evolution, we used the linear fits to the specific star formation rate-stellar mass relation for the 
three lowest redshift bins from \citet{Lara-Lopez:2013aa} for normalization. The top panel in Fig.~\ref{fig:sfrmass}
shows the original data points, which are colored based to their redshift bin, together with the main sequence relations from 
\citet[][equation (9) and table 2]{Lara-Lopez:2013aa}. The bottom panel shows the normalized data points. All five low-redshift objects discussed here
were included in the lowest redshift bin.

The five objects show a variety of star formation properties, extending 
from the main sequence of star forming galaxies down to lower specific star formation rates 
intermediate to the main sequence and the location of red quiescent galaxies. The general range of star formation properties 
covered by the NLS1s studied here is very similar to the range covered
by the type-1 QSO sample. However, none of the five NLS1s is located above the main sequence.
The highest star formation rate is found for \object{MCG-05-01-013} ($\mathrm{SFR}=2.82\ M_\odot\ \mathrm{yr^{-1}}$), which also shows one of the highest
 specific star formation rates in the sample. \object{MCG-05-01-013} is likely to be in a recent phase of active star formation and is located on the main 
 sequence of SDSS star forming galaxies in Fig.~\ref{fig:sfrmass}. 
 \object{ESO~399-IG20}, \object{MS 22549-3712}, 
 \object{TON S180},  and \object{WPVS 007} have lower H$\alpha$ star formation rates of $0.3$ to $0.8\  M_\odot\ \mathrm{yr^{-1}}$. 
 Despite a moderate star formation rate, 
 \object{WPVS 007} displays the highest specific star formation rate of the five objects. 
 The stellar mass of \object{WPVS 007} is by almost 1-2 orders of magnitude lower than the stellar masses found for the other four objects or the type-1 QSO sample.
 Nevertheless, in terms of specific star formation rate, \object{WPVS 007} is still consistent with the bulk of low-mass SDSS star forming galaxies.   

\subsection{Oxygen abundance}

\begin{figure}
\figurenum{6}
\plotone{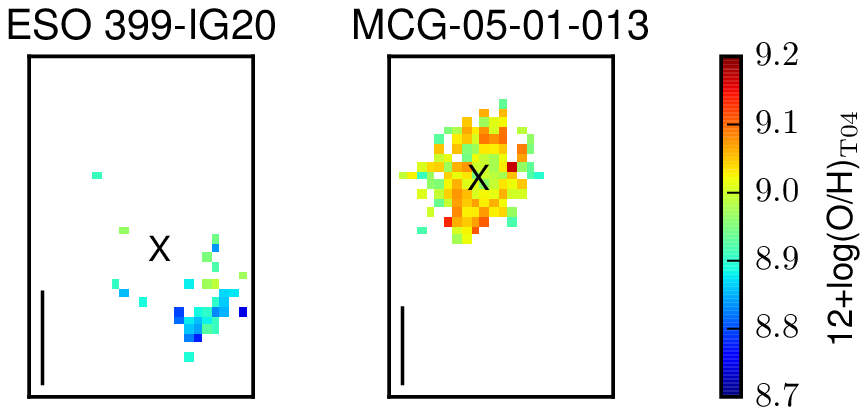}
  \caption{Maps of the oxygen abundance for the two objects with unbinned data. The oxygen abundance is derived via the
  O3N2 method \citep{Pettini:2004aa}, as described in the text. For this plot, the abundance has been converted to the T04 \citep{Tremonti:2004aa} calibration 
  using the conversion formula provided by \citet{Kewley:2008aa}. The scale bars in the maps indicate 5~kpc. The black cross shows 
  the position of the AGN, as determined from an elliptical Gaussian fit to the host-cleaned AGN cube
    integrated over the [\ion{O}{3}]$\lambda 5007$ line.}
  \label{fig:metals}
\end{figure}
\begin{figure*}
\figurenum{7}
\plotone{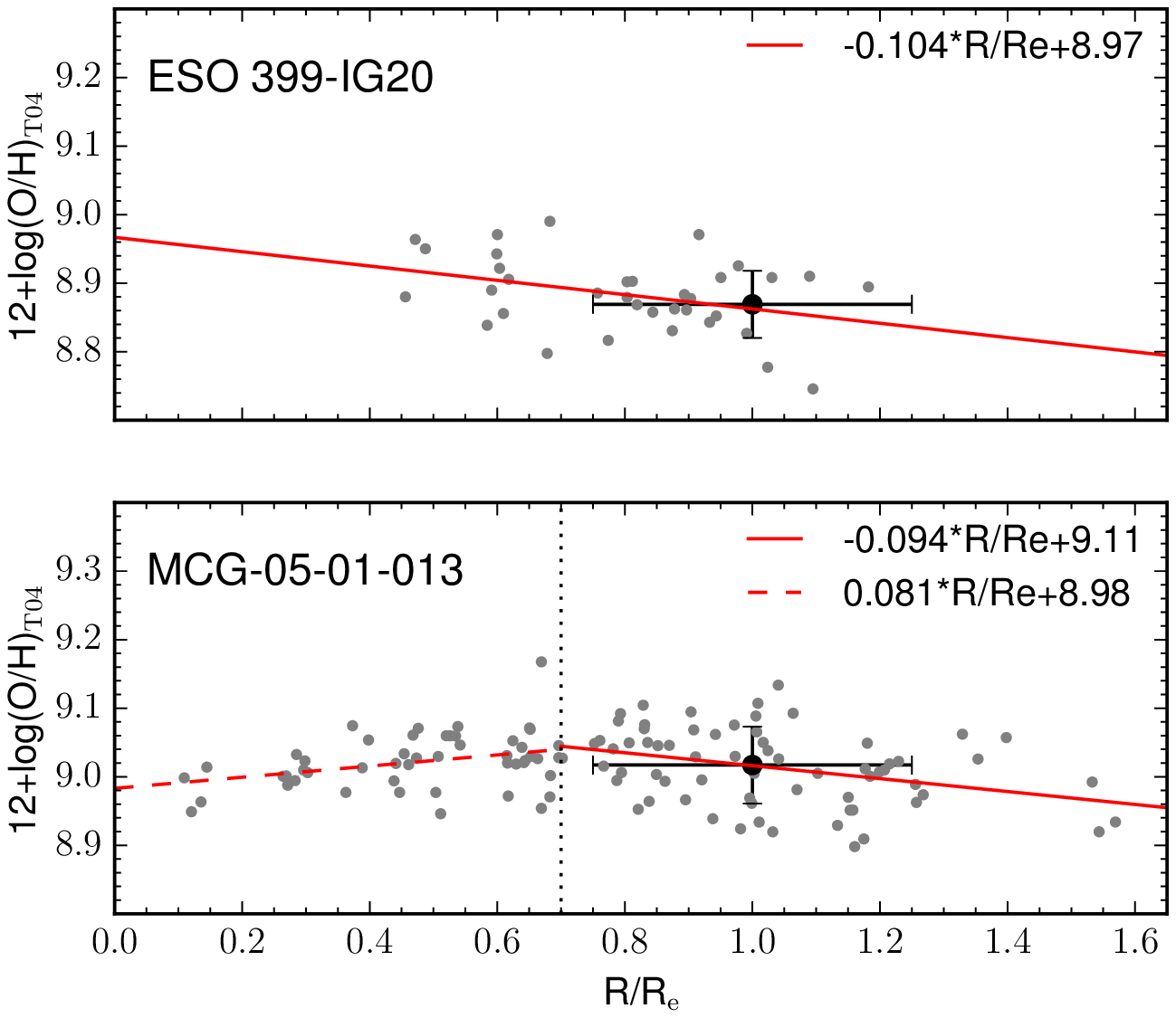}
  \caption{Radial metallicity gradients for the two oxygen abundance maps shown in Fig.~\ref{fig:metals}. The oxygen abundance values of all 
  individual pixels are plotted as grey dots. The red line is a simple linear regression, extrapolated over the full radial range shown in the plots. The parameters of the 
  linear regression are listed in the legend. For \object{MCG-05-01-013}, the linear regression has been computed separately for distances 
  $R<0.7R_\mathrm{e}$ and $R>0.7R_\mathrm{e}$ as indicated by the dashed and solid red line, respectively.
  The black error bars show the mean and standard deviation in a bin from $0.75R_\mathrm{e}$ to $1.25R_\mathrm{e}$. The gradients are not corrected for inclination.
  }
  \label{fig:metalsgrad}
\end{figure*}

The oxygen abundances and metallicity gradients for the host galaxies of the two objects with unbinned data - \object{ESO~399-IG20}  and \object{MCG-05-01-013} - 
are shown in Figs~\ref{fig:metals} and \ref{fig:metalsgrad}. The 
QSO-host deblending technique is most reliable in the wavelength region close to the lines used to probe the PSF (i.e. H$\alpha$ and H$\beta$). We use the 
 [\ion{O}{3}]$\lambda 5007$, $\mathrm{H\beta}$, [\ion{N}{2}]$\lambda 6583$, and $\mathrm{H\alpha}$ lines, in order to derive oxygen abundances based on the
 O3N2 method \citep{Pettini:2004aa}:
\begin{equation}
12+\log(\mathrm{O/H})_\mathrm{O3N2} = 8.73-0.32 \log\left(\frac{[\mathrm{OIII}]\lambda 5007/\mathrm{H\beta}}{[\mathrm{NII}]\lambda 6583/\mathrm{H\alpha}}\right).
\end{equation}
As there is evidence that the emission-line regions in the two host galaxies
are largely dominated by star formation (see Section~\ref{sec:host}), 
the oxygen abundance is calculated for all unmasked spatial pixels. 
As in \citet{Husemann:2014aa}, the oxygen abundance is converted to the T04 \citep{Tremonti:2004aa} calibration using
\begin{eqnarray}
12+\log(\mathrm{O/H})_\mathrm{T04} &=& -738.1193+258.96730x     \\ \nonumber
                                                              &  &-30.057050x^2+1.1679370x^3
\end{eqnarray}
with $x=12+\log(\mathrm{O/H})_\mathrm{O3N2}$ \citep{Kewley:2008aa}.

As can be seen in 
Figs~\ref{fig:metals} and ~\ref{fig:metalsgrad}, both of the two well-resolved 
sources show indications of radial metallicity gradients. In the outer parts of both host galaxy disks,
the oxygen abundance decreases with 
increasing galactocentric distance.  For reasons of comparison, the radius in Fig.~\ref{fig:metalsgrad} is shown in units
of effective radius, for which we use the effective radii determined from the surface-brightness fits described in Sect.~\ref{sec:deblend}.

\object{MCG-05-01-013} has a smooth metallicity gradient, with decreasing abundance towards the outer part of the host galaxy at large galactocentric distance ($\gtrsim 0.7R_\mathrm{e}$). 
Closer to the center, at R $\lesssim 0.7R_\mathrm{e}$, the gradient is
flat or even inverted. 
When using only data at 
R$\gtrsim 0.7R_\mathrm{e}$, the abundance gradient of \object{MCG-05-01-013} is $-0.094\ \mathrm{dex}/R_\mathrm{e}$, based on a linear regression.
The limited data for \object{ESO~399-IG20} only probe a range between 
about 0.45 and 1.2~$R_\mathrm{e}$, but show similar indications of decreasing abundance towards the outer parts of the
 host galaxy. The corresponding metallicity gradient is found to be $-0.1\ \mathrm{dex}/R_\mathrm{e}$.
Both gradients are very similar to the characteristic gradient of $-0.1\ \mathrm{dex}/R_\mathrm{e}$  
displayed by non-interacting galaxy disks in the local Universe 
\citep{Sanchez:2014aa}. A flattening of the oxygen abundance gradient in the inner regions has been
found in galaxy disks with circum-nuclear star forming rings \citep{Sanchez:2014aa}. Such a star forming ring may be associated with the 
flattening of the metallicity gradient in the central part of the host galaxy of \object{MCG-05-01-013}.

The oxygen abundances for the two well-resolved sources, measured at $R_\mathrm{e}$ and extrapolated to the centre, are listed in Table~\ref{table:metal}.
The oxygen abundance at $R_\mathrm{e}$ is derived from the mean and standard deviation of all
available abundance data between $0.75R_\mathrm{e}$ and $1.25R_\mathrm{e}$. This is the same method as used for the 
 $z<0.2$ type-1 QSOs by \citet{Husemann:2014aa}, which we adopt for reasons of comparison. 
The central oxygen abundance shown in Table~\ref{table:metal} is extrapolated from the value at $R_\mathrm{e}$ using a
metallicity gradient of $-0.1\ \mathrm{dex}/R_\mathrm{e}$. 
For \object{MCG-05-01-013}, we also show the central oxygen abundance resulting 
from the linear regression for $R<0.7R_\mathrm{e}$, i.e
taking into account the central flattening of the metallicity gradient.

\floattable
\begin{deluxetable}{lccc}
\tablecaption{Oxygen abundance \label{table:metal}   }
\tablecolumns{4}
\tablewidth{0pt}
\tablehead{ 
\colhead{Object} & 
\colhead{$R_\mathrm{e}$} & 
\colhead{$12+\log(\mathrm{O/H})_\mathrm{T04}$ } & 
\colhead{$12+\log(\mathrm{O/H})_\mathrm{T04}$}  \\ 
\colhead{}          &  
\colhead{(arcsec)} &  
\colhead{at $R_\mathrm{e}$ } & 
\colhead{at $R_0$  }  }
\decimalcolnumbers
\startdata
 \object{ESO~399-IG20}    & 10.1  & $8.87\pm0.05$ & $8.97\tablenotemark{a}\pm0.05$ \\ 
  \object{MCG-05-01-013}  &  5.6 & $9.02\pm0.06$ & $9.12\tablenotemark{a}\pm0.06$ $\slash$ $8.98\tablenotemark{b} $ \\
 \enddata
\tablecomments{Column 1: Object name; Column 2: Effective radius $R_\mathrm{e}$; 
Column 3: Average oxygen abundance at $R_\mathrm{e}$; 
Column 4: Extrapolated central oxygen abundance.}
\tablenotetext{a}{Extrapolated from $R_\mathrm{e}$ assuming -0.1~dex/$R_\mathrm{e}$}
\tablenotetext{b}{Extrapolated via linear regression using the flat gradient found for $R<0.7R_\mathrm{e}$.}
\end{deluxetable}  
  
\begin{figure}
\figurenum{8}
\epsscale{0.85}
\plotone{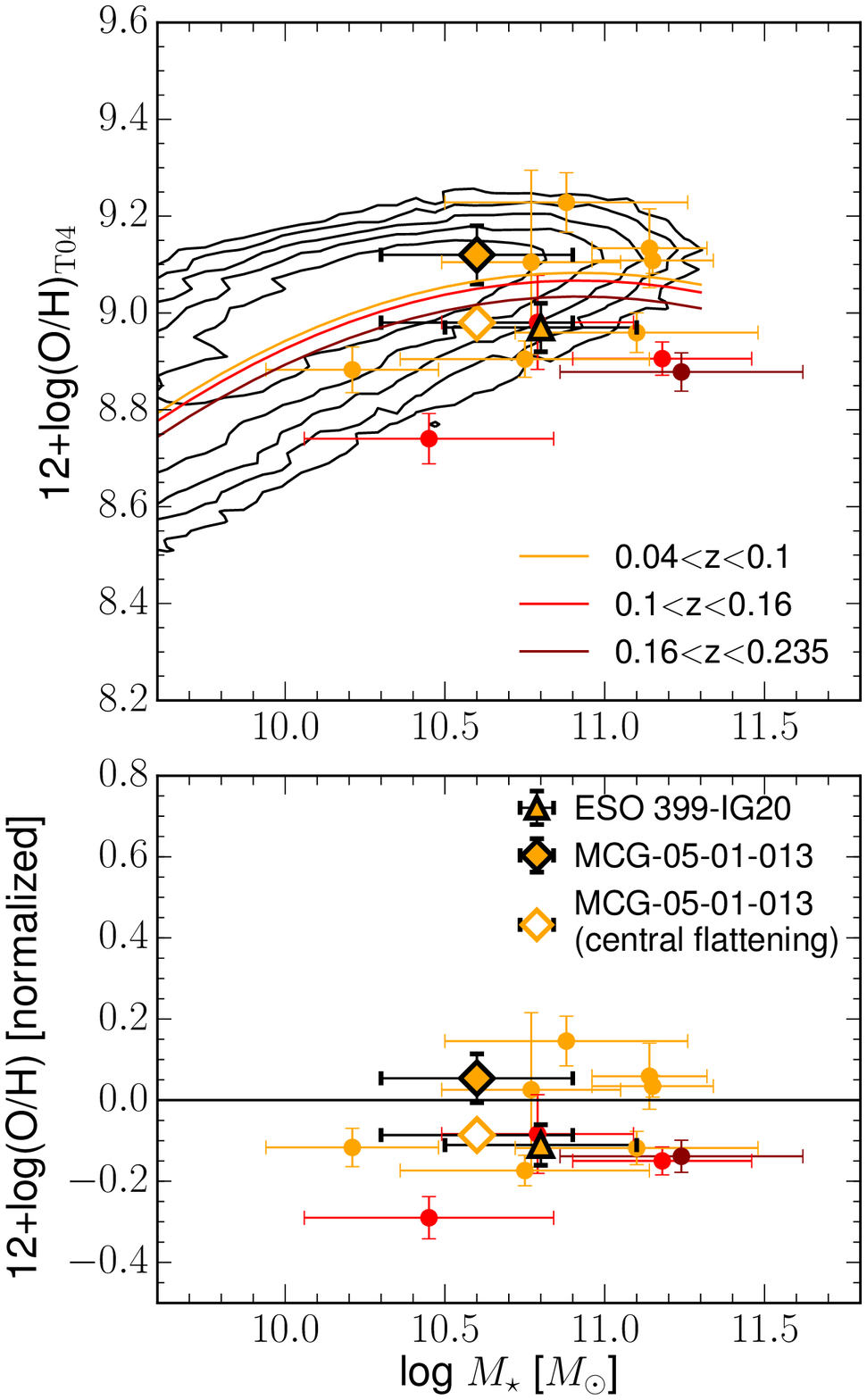}
  \caption{Comparison of the metallicity measurements for \object{ESO~399-IG20}  and \object{MCG-05-01-013} with SDSS galaxies, the mass-metallicity relations for the three
  lowest redshift bins derived by \citet{Lara-Lopez:2013aa}, and $z<0.2$ type-1 QSOs from \citet{Husemann:2014aa}. 
  The color of the data points for \object{ESO~399-IG20}, \object{MCG-05-01-013}, and the type-1 QSOs indicates their corresponding redshift bin.
  The two NLS1s are shown with large symbols, as described in the plot legend. The type-1 QSOs are marked by filled circles. The metallicity shown
  for the two NLS1s and the type-1 QSOs corresponds to the central oxygen abundance, as extrapolated from
   the abundance at $R_\mathrm{e}$ assuming a metallicity gradient of -0.1~dex/$R_\mathrm{e}$. The open diamond symbol for 
   \object{MCG-05-01-013} shows the central oxygen abundance corresponding to the flat gradient found for the central region of this galaxy.
   Top panel: Direct comparison of the abundance data with the SDSS galaxies from the SDSS MPA/JHU galaxy catalogue 
   \citep{Brinchmann:2004aa} (contours) and the fits to the mass-metallicity relation in the three lowest redshift bins presented by \citet{Lara-Lopez:2013aa} (solid lines).
   Bottom panel: The NLS1 and type-1 QSO metallicities after normalization using the fits to the mass-metallicity relation from \citet{Lara-Lopez:2013aa} for the corresponding redshift bin.
  \label{fig:OHmass}}
\end{figure}
 In Fig.~\ref{fig:OHmass} (top panel), the two sources with measurements for the central oxygen abundance are plotted in the 
mass-metallicity plane and compared to SDSS galaxies and the type-1 QSOs from \citet{Husemann:2014aa}. In addition, the plot shows fits to the mass-metallicity relation for the three lowest redshift bins presented by \citet{Lara-Lopez:2013aa}. The data points are colored based on their redshift bin. 
In the bottom panel, the same data points are compared after normalization based on the fitted 
mass-metallicity relations from \citet{Lara-Lopez:2013aa}. 
The central oxygen abundances for \object{ESO~399-IG20} and \object{MCG-05-01-013} are about 0.1~dex below
the low-redshift mass-metallicity relation from \citet{Lara-Lopez:2013aa}. 
(For \object{MCG-05-01-013}, this statement is true when considering the flat/inverted central 
metallicity gradient, resulting in the central abundance marked by the open diamond.) In their type-1 QSO sample,
\citet{Husemann:2014aa} found a trend of bulge-dominated hosts to preferentially lie at the lower end or below the mass-metallicity relation, while disk-dominated
hosts mostly follow the main relation. In comparison to the type-1 QSO sample, the two objects analyzed here seem to occupy an intermediate region between the 
bulge- and disk-dominated type-1 QSO host galaxies. 
 
\subsection{Notes on individual objects}

\subsubsection{ESO 399-IG20}

\object{ESO 399-IG20} belongs to a galaxy pair at $z=0.025$. 
An optical long-slit spectrum of \object{ESO 399-IG20} was analyzed by \citet{Dietrich:2005aa} as part of a study of the ionizing continuum of NLS1s. 
Among our five objects, \object{ESO 399-IG20} shows the second largest BH mass, second lowest Eddington ratio, and second weakest \ion{Fe}{2}
emission after \object{MCG-05-01-013} and, therefore, rather mild NLS1 characteristics.
For the broad H$\beta$ component, we find an FWHM of $1900\ \mathrm{km\ s^{-1}}$ (Table~\ref{table:AGNpars}), which is just below the classical 
NLS1 criterion of FWHM(H$\beta) < 2000\ \mathrm{km\ s^{-1}}$.
The two Gaussians used to fit the broad H$\beta$ component in \citet{Dietrich:2005aa}, ($\mathrm{FWHM(H\beta\ broad) =2425\pm 121\ km\ s^{-1}}$ and 
($\mathrm{FWHM(H\beta\ intermediate) = 944\pm 47\ km\ s^{-1}}$), are about 1.8 and 1.6 times narrower than 
the very broad and broad components found here (Tables~\ref{table:fluxes}), 
while the narrow component reported by \citet{Dietrich:2005aa}, ($\mathrm{FWHM(H\beta\ narrow) = 479\pm 24\ km\ s^{-1}}$), is about 1.4 times broader than the narrow component resulting from our fit. Differences in the line width measurements might result from the AGN-host deblending. \citet{Pozo-Nunez:2013aa} 
studied photometric reverberation mapping of \object{ESO 399-IG20} and obtained an interpolated host-subtracted AGN flux density
at 5100~\AA\ of $f_{5100} = (2.66\pm0.22)\times 10^{-15}\ \mathrm{erg\ s^{-1}\ cm^{-2}\ \AA^{-1}}$, which can be compared with the corresponding flux density of
$f_{5100} = (1.8\pm 0.2) \times 10^{-15}\ \mathrm{erg\ s^{-1}\ cm^{-2}\ \AA^{-1}}$ found here. Based on their analysis, \citet{Pozo-Nunez:2013aa} suggest a disk-like BLR geometry seen nearly face-on. 
\citet{Dietrich:2005aa} find a large host-galaxy contribution of 51\% to the total 5100~\AA\ flux and suggest
an S0-type host. The WiFeS data cube presented here covers a small part of the western companion galaxy 
(Fig.~\ref{fig:maps}a). The companion galaxy is characterized by blueshifted velocities
($\sim -190$ to $-230\ \mathrm{km\ s^{-1}}$) with respect to the range of gas velocities found for \object{ESO 399-IG20}
($\sim -190$ to $190\ \mathrm{km\ s^{-1}}$).
In how far the velocity field of \object{ESO 399-IG20} is perturbed by the interaction is difficult to judge based on the current data set. 
The line ratios in the AGN-subtracted host galaxy spectra of \object{ESO 399-IG20} largely indicate
ionization by star formation as the dominant mechanism. The pixels associated with the companion galaxy, shown as red points in the bottom left panel of Fig.~\ref{fig:maps}a,
reach slightly higher [\ion{O}{3}]$\lambda 5007$/H$\beta$
ratios, which may be related to a contribution from AGN ionization.
\object{ESO 399-IG20} belongs to the three sources of this study that show specific star formation rates
intermediate to the main sequence and the location of red quiescent galaxies (Fig.~\ref{fig:sfrmass}).
The central oxygen abundance is just below the low-redshift mass-metallicity relation (Fig.~\ref{fig:OHmass}).

\subsubsection{MCG-05-01-013}
\object{MCG-05-01-013} is part of an overlapping pair of galaxies that are not physically connected. \object{MCG-05-01-013} is an 
Sb-type galaxy at $z=0.031$, while the galaxy to the west (Fig.~\ref{fig:maps}b) is a background galaxy at a redshift of $z=0.062$ (data taken from
NASA/IPAC Extragalactic Database), as also confirmed by our new spectra. 
As part of the soft X-ray selected AGN sample discussed by \citet{Grupe:2004ab} and \citet{Grupe:2010aa}, \object{MCG-05-01-013} 
shows $\mathrm{FWHM(H\beta)=2400\pm 200\ km\ s^{-1}}$, based on which the broad H$\beta$ component of \object{MCG-05-01-013} 
just exceeds the classical NLS1 definition of $\mathrm{FWHM(H\beta) = 2000\ km\ s^{-1}}$. 
Our analysis suggests a smaller H$\beta$ width of $1850\pm 200\ \mathrm{km\ s^{-1}}$, which matches the classical NLS1 line width
criterion. \citet{Grupe:2004ab} find \ion{Fe}{2}/H$\beta=0.81$, which is larger by a factor of $\sim 1.8$ than the value derived here, and
FWHM([\ion{O}{3}])$=250\pm 100\ \mathrm{km\ s^{-1}}$, which is comparable to our results (Table~\ref{table:fluxes}). 
The bolometric luminosity of $L_\mathrm{bol}=10^{36.87}$ W, (or $L_\mathrm{bol}=10^{43.87}$ $\mathrm{erg\ s^{-1}}$), 
derived from the optical/UV and X-ray spectrum by \citet{Grupe:2004ab} is very similar to the bolometric luminosity 
of $L_\mathrm{bol}=10^{43.65}$ $\mathrm{erg\ s^{-1}}$ derived from the 5100~\AA\ continuum flux in this paper.
Based on our analysis, \object{MCG-05-01-013}
shows the lowest Eddington ratio and weakest
\ion{Fe}{2} emission of the five objects (Tables~\ref{table:fluxes} and \ref{table:AGNpars}). This suggests that it
might represent the least prototypical NLS1 characteristics for the purpose of the present paper.
The host galaxy is characterized by a rather regular
gas rotation pattern suggesting disk rotation. The H$\alpha$ emission (Fig.~\ref{fig:maps}b) shows a bar-like structure extending over the inner $\sim 5\ \mathrm{kpc}$ with possibly two emerging spiral arms and a prominent H$\alpha$ knot to the south of the nucleus. Gas excitation in the host galaxy is dominated by star formation.
\object{MCG-05-01-013} has the highest star formation rate in our sample and a specific
star formation rate consistent with the main sequence of star forming galaxies (Fig.~\ref{fig:sfrmass}). The host galaxy shows a clear metallicity gradient with
central flattening, based on which the central oxygen abundance of \object{MCG-05-01-013} is just below the low-redshift mass-metallicity relation.

\subsubsection{MS 22549-3712}
\object{MS 22549-3712} at $z=0.039$ is a disk galaxy with a likely pseudo-bulge \citep{Mathur:2012aa}. 
The gas velocity field in Fig.~\ref{fig:maps}c shows a clear velocity gradient in agreement with disk rotation. 
The line ratios for the host galaxy lie below the theoretical maximum starburst
line from \citet{Kewley:2001aa} (Fig.~\ref{fig:maps}c) and are most likely dominated by ionization from star formation, though a small contribution from 
AGN or shock ionization may be present. Possible evidence for redshifted [\ion{S}{2}] emission from the unresolved NLR 
(Table \ref{table:fluxes}) could also suggest shocks, but this interpretation remains speculative. The star formation rate
derived for \object{MS 22549-3712} is rather low ($0.5\pm0.4\ \mathrm{M_\odot\ yr^{-1}}$) and corresponds to a specific star formation rate
intermediate to the main sequence of star forming galaxies and red quiescent galaxies in Fig.~\ref{fig:sfrmass}. 
\citet{Grupe:2004ab} and \citet{Grupe:2010aa} report a broad H$\beta$ width of $1530\ \mathrm{km\ s^{-1}}$,
FWHM([\ion{O}{3}])$=510\pm 60\ \mathrm{km\ s^{-1}}$, \ion{Fe}{2}/H$\beta=0.53$ and $\log \lambda L_\mathrm{5100}=36.34\ \mathrm{W}=43.34\ \mathrm{erg\ s^{-1}}$. 
These values are similar to the ones derived from our host-deblended nuclear
spectrum (Tables~\ref{table:fluxes} and \ref{table:AGNpars}).
The nuclear spectrum of \object{MS 22549-3712} indicates an average Eddington ratio and BH mass compared to the other four sources discussed here.

\subsubsection{TON S180}

\object{TON S180} at $z=0.062$ is the highest redshift source of our sample. It shows the highest Eddington ratio
and strongest \ion{Fe}{2} emission (Tables~\ref{table:fluxes} and \ref{table:AGNpars}). \object{TON S180} is a well-studied object and frequently included in NLS1 samples.
\citet{Grupe:2004ab} report FWHM([\ion{O}{3}])$=630\pm 60\ \mathrm{km\ s^{-1}}$ and \ion{Fe}{2}/H$\beta$=0.9. 
These values as well as the $\log L_\mathrm{bol}$ and $\log \lambda L_\mathrm{5100}$ listed by \citet{Grupe:2004ab} are roughly
in agreement with the results from our fitting method (Table~\ref{table:fluxes} and Table~\ref{table:AGNpars}). 
\citet{Grupe:1999aa} measure a broad H$\beta$ width of FWHM(H$\beta$)$=1380\pm100\ \mathrm{km\ s^{-1}}$. A slightly smaller width 
of about $1080\ \mathrm{km\ s^{-1}}$ is
reported by \citet{Wang:2001aa}, \citet{Marziani:2003ab}, and \citet{Grupe:2004ab}. This width is consistent with the 
FWHM(H$\beta$)$=1060\pm50\ \mathrm{km\ s^{-1}}$ found here (Table~\ref{table:AGNpars}).
The host galaxy of \object{TON S180} is a disk galaxy with spiral arms  and a likely classical bulge \citep{Mathur:2012aa}. 
The gas velocity maps in Fig.~\ref{fig:maps}d show a velocity gradient across the nucleus, consistent
with disk rotation. The signal-to-noise ratio of the current data set is not sufficient for a line-ratio analysis even after binning. Assuming that the
observed host galaxy H$\alpha$ flux is indicative of star formation, we derive a moderate star formation rate and a specific star formation rate 
that is intermediate 
to that of main sequence galaxies and quiescent red galaxies.

\subsubsection{WPVS 007}

\object{WPVS 007} at $z=0.029$ is a rare example of a nearby low-luminosity AGN which developed a broad-absorption line outflow
between 1996 and 2003 \citep{Leighly:2009aa}. This NLS1 is known 
to show 
strong variability in X-rays and in the UV, likely driven by time-variable absorption \citep{Grupe:2013aa}.
While the integrated nuclear WiFeS spectrum for \object{WPVS 007} was presented in \citet{Grupe:2013aa}, the present paper shows 
a full analysis of the nuclear spectrum and host galaxy data cube after AGN-host separation.
\object{WPVS 007}
has the second narrowest H$\beta$ line width, smallest BH mass, and second 
highest Eddington ratio of the five objects (Table~\ref{table:AGNpars}). The BH mass estimated from the H$\beta$ line width in this study
is roughly consistent with the value  
of $M_\mathrm{BH}=10^{6.6}\ M_\sun$ derived from the
near-infrared Pa$\alpha$ line by \citet{Busch:2016aa}. 
The host galaxy is compact, shows clear evidence for 
a gas velocity gradient across the nucleus (Fig.~\ref{fig:maps}e), and has the lowest stellar mass of the five objects (Table~\ref{table:sfr}). 
\citet{Busch:2014aa} suggest a 
largely bulge-dominated host galaxy, with possible indications of sub-structure and a pseudo-bulge. 
The line ratios in Fig.~\ref{fig:maps}e are generally consistent with ionization by star forming regions. 
There is marginal evidence for a higher
 [\ion{O}{3}]/H$\beta$ ratio in 1-2 of the binned pixels to the north of the nucleus, which may indicate
some contribution from AGN or shock ionization. Based on the integrated host galaxy H$\alpha$ emission, \object{WPVS 007} has the 
lowest star formation rate of all five objects. In terms of specific star formation rate, \object{WPVS 007} lies just below the main sequence of low-redshift
star forming galaxies.

\section{Summary and Conclusions}\label{sec:conclusion}
In how far the nuclear characteristics of NLS1-type AGN are reflected in 
star formation rates, gas metallicity, or gas ionization properties on host galaxy scales is still a matter of debate. 
In this paper we have presented optical integral field spectroscopy of five NLS1s addressing this question in the form of an initial case study. 

Based on the host-deblended AGN spectra, all five sources are consistent with the classical definition of NLS1s, i.e. $\mathrm{FWHM(H\beta)} < 2000\ \mathrm{km\ s^{-1}}$ and [\ion{O}{3}]/$\mathrm{H\beta} < 3$. In agreement with previous findings for narrow-line AGN, the line profiles of the broad H$\alpha$ and H$\beta$ components
of the five sources are non-Gaussian with an FWHM/$\sigma$ ratio close to 1. The BH masses, Eddington ratios, and \ion{Fe}{2} strengths
overlap with the corresponding distributions shown by the large SDSS-based NLS1 sample studied by \citet{Rakshit:2017aa}.
Given that NLS1s are unlikely to form a distinct class but rather a smooth transition towards one extreme end of the AGN parameter space, it is not unexpected
to find a variety of NLS1-type properties in this case study. In this respect, 
 \object{TON S180}, \object{WPVS 007}, and \object{MS 22549-3712} show the more prototypical NLS1 properties, while 
\object{ESO~399-IG20} and \object{MCG-05-01-013} have weaker \ion{Fe}{2} emission and lower Eddington ratios.

Taking advantage of the AGN-deblended host galaxy data, we have measured gas ionization and kinematics of the five objects and analyzed star formation rates and host gas metallicity for the sources with sufficient data quality.
The gas kinematics for all five sources shows evidence of rotation. In combination with literature data and our $g-r$ colors, we conclude that 
 \object{TON S180}, \object{MS 22549-3712}, \object{MCG-05-01-013} have disk-type host galaxies, while the host types for \object{ESO~399-IG20} and \object{WPVS 007} are more uncertain. It is therefore possible that secular evolution plays an important role in at least some of the five objects.

At the spatial scales of 2-3~kpc probed by our study, there is no clear evidence of 
any significant AGN contribution to the gas ionization in the host galaxies. Instead, the predominant ionization mechanism at these scales appears to be star formation. 
This does not rule out any AGN-ionized extended NLRs on smaller scales.
For the [\ion{O}{3}] luminosities of the five sources, the size-luminosity relation for extended NLRs would predict NLR sizes that are 2-7 times smaller than the spatial resolution
achieved here
\citep[e.g.][]{Bennert:2002aa, Liu:2013aa,Hainline:2014aa,Husemann:2014aa}. A systematic analysis of NLS1s
in view of the size-luminosity relation is still missing. Initial studies find that extended emission-line regions are preferentially detected around
QSOs with a large H$\beta$ width and small \ion{Fe}{2} equivalent width, while they remain undetected around QSOs with 
a small H$\beta$ width and large \ion{Fe}{2} equivalent width \citep{Husemann:2008aa}. It has also been proposed that extended emission-line regions 
correlate with BLR metallicity, in the sense that
luminous extended emission-line regions are rather found for objects with low BLR metallicities \citep{Fu:2007aa}. 
As NLS1s might be characterized by unusually high BLR metallicities \citep[e.g.][]{Shemmer:2002aa, Fields:2005aa}, 
this could imply that extended AGN ionization around NLS1 nuclei is compact or absent. Higher spatial resolution IFU data for NLS1s will be needed to
further investigate the properties of extended NLRs in NLS1s in light of the size-luminosity relation.

Using the total host galaxy H$\alpha$ luminosities, we derived integrated star formation rates.
The integrated star formation rates for four of the sources are $0.3$ to $0.8\ \mathrm{M_\odot\ yr^{-1}}$, while 
\object{MCG-05-01-013} shows a higher rate of $2.82\ \mathrm{M_\odot\ yr^{-1}}$.
It has been hypothesized that NLS1-type AGN are hosted in gas-rich rejuvenated galaxies with enhanced star formation activity, 
because high BH accretion rates might only be achieved with the presence of sufficiently large gas reservoirs \citep[e.g.][]{Mathur:2000aa}.
The few systematic studies focussing on star formation activity in larger samples of NLS1s seem to support this scenario.
Based on \textit{Spitzer} mid-infrared spectroscopy of a sample of $z<0.02$ NLS1s and BLS1s, \citet{Sani:2010aa} 
find enhanced circum-nuclear star formation activity in NLS1s. 
The small size of the present sample prohibits any statistical conclusions. But it is worth noting that all objects lie below the
main sequence of star forming galaxies, except for \object{MCG-05-01-013} which we consider the least prototypical representative of NLS1 characteristics in 
this small sample.

The host oxygen abundance was studied for the two sources 
with sufficient data quality (\object{ESO~399-IG20} and \object{MCG-05-01-013}). Neither of these shows evidence of
unusually high metallicities. They rather overlap with the lower-metallicity portion of the low-redshift mass-metallicity relation. 
On nuclear scales, 
NLS1s have been reported to show unusually high \ion{N}{5}/\ion{C}{4} ratios, which has led to the suggestion that either NLS1s have
unusually high BLR metallicities comparable to those of luminous high-redshift quasars \citep{Hamann:1993aa,Nagao:2006aa}, 
or their \ion{N}{5}/\ion{C}{4} ratios are not a reliable abundance indicator \citep{Shemmer:2002aa}. 
High BLR metallicities in NLS1s could be a manifestation of a general correlation between BLR metallicity and Eddington ratio \citep[e.g.][]{Shemmer:2004aa,Shin:2013aa}, though the existence
of such a correlation remains controversial \citep{Du:2014aa}.
\citet{Fields:2005aa} propose that the high metallicities in NLS1s could be associated with a different chemical enrichment process, 
resulting in N/C ratios that do not scale with metallicity $\propto Z^2$ as commonly assumed in 
theoretical models. 
Extending similar analysis methods as the ones presented here to larger samples of NLS1s can probe whether the unusual BLR metallicities are reflected in
any special characteristics on host galaxy scales.

The results presented in this paper demonstrate the potential of using optical integral field spectroscopy together
with an AGN-host decomposition for a comparison of nuclear and host galaxy parameters in NLS1s. Extended studies of this kind, e.g. based on MUSE (VLT) observations, are likely to further
improve our understanding of objects occupying the NLS1 region of the AGN parameter space.

\acknowledgements
The authors thank the anonymous referee for a comprehensive review of the manuscript and valuable comments.
J.S. is grateful for support under the ESO scientific visitor program for a five-week visit to ESO Garching in August/September 2014, during which large parts
of the data analysis for this paper were completed. For parts of this work, J.S. also acknowledges the European Research Council for the
Advanced Grant Program Num 267399-Momentum. G.B. acknowledges support from the Collaborative Research Centre 956, sub-project A2, and the Bonn-Cologne Graduate School of Physics and Astronomy, both funded by the Deutsche Forschungsgemeinschaft (DFG). S.K. acknowledges support from NSFC
  grant 11273027. 
This research has made use of the NASA/IPAC Extragalactic Database (NED) which is operated by the Jet Propulsion Laboratory, California Institute of Technology, under contract with the National Aeronautics and Space Administration.

\facility{ATT}

\bibliography{NLS1bib.bib}



\listofchanges

\end{document}